%% file: ms.tex
\documentclass[apj]{emulateapj}
\usepackage{amsmath}
\usepackage{apjfonts}
\usepackage{multirow}
\usepackage{hyperref}
\usepackage{array}
\usepackage{enumerate}
\usepackage{appendix}
\input{Definitions.tex}

\newcommand{\sigT}{\sigma_{\rm T}}

\begin{document} \title{Numeric spectral radiation hydrodynamic calculations of supernova shock breakouts}
\author{Nir Sapir and Dorri Halbertal}
\affiliation{Dept. of Particle Phys. \& Astrophys., Weizmann Institute of Science, Rehovot 76100, Israel}

\begin{abstract}
We present here an efficient numerical scheme for solving the non-relativistic 1D radiation-hydrodynamics equations including inelastic Compton scattering, which is not included in most codes and is crucial for solving problems such as shock breakout. The devised code is applied to the problems of a steady-state planar radiation mediated shock (RMS) and RMS breakout from a stellar envelope. The results are in agreement with those of a previous work on shock breakout \citep{Sapir13}, in which Compton equilibrium between matter and radiation was assumed and the "effective photon" approximation was used to describe the radiation spectrum. In particular, we show that the luminosity and its temporal dependence, the peak temperature at breakout, and the universal shape of the spectral fluence derived in this earlier work are all accurate. Although there is a discrepancy between the spectral calculations and the effective photon approximation due to the inaccuracy of the effective photon approximation estimate of the effective photon production rate, which grows with lower densities and higher velocities, the difference in peak temperature reaches only $30\%$ for the most discrepant cases of fast shocks in blue supergiants. The incompatibility of the stellar envelope shock breakout model results with observed properties of X-ray flashes and the discrepancy between the predicted and observed rates of X-ray flashes \citep{Sapir13} remain unexplained.
\end{abstract}

\section{Introduction}\label{sec:Introduction}

X-ray flashes (XRFs) and low luminosity gamma-ray bursts (LLGRBs) have been observed in recent years to precede in several cases the optical emission of supernova (SN) explosions \citep{Campana06,Woosley06,Soderberg08,Cano11,Starling11,Melandri12}.
XRFs with characteristic photon energies in the range of 5 to $50 \keV$ have been interpreted as SN shock breakouts with non-relativistic or mildly relativistic velocities ($v/c \lesssim 0.5$), either from the stellar envelope or possibly from an extended stellar wind.
This interpretation is based mainly on the total bolometric energy emitted in these XRFs, $10^{46}-10^{49} \erg$, which is compatible with the shock breakout scenario, and on the emission time scale, $\sim 300-3000 \se$, which is much shorter than the typical rise time of the SN optical emission and might be compatible with the light crossing time of the shock breakout radius.
However, the predicted emission spectrum and color temperature vary between different studies, and are essential for the identification of XRFs with the shock breakout phase of SNe.

Radiation-hydrodynamics codes usually apply the diffusion approximation to solve the radiation transport either in the zeroth or first angular moment expansion.
The main difference between the different codes lies in the details of the radiation-matter interaction.
Different works in the literature have reported on numerical calculations of SN shock breakout from a stellar envelope, where both the bolometric and the spectral characteristics of the burst were evaluated for specific progenitors \citep[e.g.][]{Grassberg71,Imshennik77,Falk78,Klein78,Lasher79,Ensman92,Blinnikov98,Blinnikov00,Tominaga11,Tolstov13}.
These works considered different levels of simplifications for the radiative interactions.
In the simplest approximation, the plasma and the radiation are assumed to be in local thermal equilibrium (LTE).
In the two temperature approximation, or single group approximation, the radiation and the plasma have different temperatures and are not assumed a-priori to be in LTE.
In this formulation the radiation temperature is derived from the radiation energy density assuming a Planck spectrum and the radiation-matter coupling includes emission and absorption, and can also include inelastic Compton scattering as a bolometric heating (or cooling) term.
The absorption and inelastic scattering terms are also calculated assuming a Planck spectrum.
In the multi-group approximation the radiation spectrum is described as a function of photon energies and so absorption and inelastic scattering can be evaluated correctly for an arbitrary spectrum.
However, inelastic Compton scattering is usually neglected in codes applying this approximation.

The different approximations provide very different results for the shock breakout problem.
For example, for a SN from a red supergiant with a $15 \Msun$ mass and $R\sim 10^{13} \cm$ radius with $10^{51}\erg$ in explosion energy the breakout predictions range from emission of $\sim 10 \eV$ photons assuming LTE \citep{Lasher79} to emission of $\sim 100 \eV$ photons in two temperature approximations \citep{Falk78,Klein78,Tominaga11}.
A significant contribution of $>1 \keV$ and possibly $100 \keV$ photons was also predicted \citep{Falk78,Klein78}.
The main source for the difference in the predicted energy of the emitted photons is the use of an LTE assumption versus a single-group calculation.
In addition, the X-ray emission ($>1 \keV$) in these calculations results from a formation of a viscous shock, which was shown to be absent in the breakout of a radiation mediated shock (RMS) from a stellar envelope \citep{Lasher79,Epstein81,Sapir11}.
Similarly, breakout calculations performed for the blue supergiant progenitor of SN1987A have yielded a peak effective temperature of $\sim 20-40 \eV$ when LTE was assumed \citep{Woosley88,Shigeyama88} and a peak color temperature of $\sim 100 \eV$ in single-group or multi-group calculations that do not include inelastic Compton scattering \citep{Ensman92,Blinnikov00}.

Despite the fact that electron scattering dominates the opacity in stellar envelopes, most previous works only considered its effect on the photon mean free path and on the bolometric energy exchange with matter, and did not include inelastic Compton scattering in the calculation of the radiation spectrum \citep[with the exception of][]{Imshennik77}.
But at the fronts of RMSs propagating in stellar envelopes there is a high Compton y-parameter and a low effective optical depth for absorption, so Compton losses (gain) are important for high (low) energy photons, while thermalization is unimportant \citep{Weaver76,Katz10}.
Thus, the electrons and the radiation are able to reach Compton equilibrium, and the photons saturate in energy to roughly the matter temperature.
As these effects become significant at high shock velocities, $v>0.05 c$, Comptonization is specifically important for small progenitors, and the temperatures can be orders of magnitude higher than the LTE values.

Previous works have provided the exact bolometric light curve and the expected (approximate) emission spectrum following breakout of a planar non-relativistic RMS from a stellar envelope, using the diffusion approximation with constant opacity \citep{Sapir11,Katz12,Sapir13}.
In particular, the temperature calculations were based on the effective photon approximation, which assumes local Compton equilibrium between the matter and the radiation \citep{Weaver76,Katz10}.
In this approximation most of the radiation energy is assumed to be characterized by a Wien spectrum, and the photon production rate includes only the photons that can Comptonize with the matter, with an absorption correction.
The produced radiation spectrum is the result of Comptonization, and is defined by the photon number density and the radiation temperature.

In these works simple approximations were obtained for the time varying bolometric luminosity and surface temperature, as well as for the peak temperature and the observed spectrum.
Using the fact that the results are not sensitive to the assumed progenitor density profile and to mild asymmetries, the spectral emission following breakout from any progenitor can be estimated without the need to perform a full radiation-hydrodynamical calculation for a specific progenitor.

Although it is a justified assumption for most of the flow regimes in the breakout problem, the effective photon approximation is still an uncontrolled approximation.
It involves two photon-energy moment equations to be solved, for the bolometric energy density and for the photon number density, where the closure relation allowing the solution of the equations is the assumption of local Compton equilibrium.
Detailed calculations are required in order to fully test this assumption.
Such calculations can then be used to study cases in which no Compton equilibrium is expected (e.g., extremely low densities).
The complication of the radiation-hydrodynamics equations describing the evolution of the radiation spectrum is that they are non-local both in space and in photon energy.
Challenges of constructing a numerical code to solve these equations include maintaining energy conservation in the radiative interactions between the matter and the radiation, and maintaining numerical stability in steady-state, both without restricting the time step too much.
In particular, the exponential tails that generally characterize the spectrum at high photon energies need to be addressed by appropriate numerical techniques.

In this work, the problem of a planar non-relativistic RMS breaking out from a stellar envelope is investigated using (spectral) radiation-hydrodynamics numerical calculations.
A numerical scheme is devised for solving the radiation-hydrodynamics equations in the non-relativistic regime, describing the dominant radiative processes that couple the radiation to the matter .
In this regime the Kompaneets equation and the diffusion equation are valid approximations for describing Compton scattering and radiation transport, respectively.
Additionally, the plasma is assumed to be fully ionized and described by an ideal equation of state, and bremsstrahlung is assumed to dominate radiative emission.
We compare the results of these detailed calculations to the results of calculations with the effective photon approximation, and show that the effective photon approximation is valid and its results are affirmed.

The paper is organized as follows.
The approximations and the radiation-hydrodynamics equations describing the problem are presented in \sref{sec:The radiation-hydrodynamics equations}.
The numerical solution for the problem of a steady-state RMS is presented in \sref{sec:Radiation mediated shock in a homogenous medium}, and the numerical solution for problem of an RMS breakout from a stellar envelope is presented in \sref{sec:Planar shock breakout from a stellar surface}.
The conclusions of this work are described in \sref{sec:Discussion}.

\section{The radiation-hydrodynamics equations}\label{sec:The radiation-hydrodynamics equations}

Radiation hydrodynamics describes the motion of matter and the radiation transport, which are coupled through different interactions.
Here we consider fully ionized plasmas where the ions and the electrons move together and share the same temperature $T$.
The radiation on the other hand is not in LTE, and a spectral description is given as a function of the photon energy $\varepsilon$.
We consider the following assumptions:
\begin{itemize}
\item Non-relativistic velocities, $v \ll c$.
\item Non-relativistic energies, $\varepsilon, T \ll m_{\rm e}c^2$.
\item Planar geometry.
\item Locally isotropic radiation field.
\end{itemize}
These assumptions greatly simplify the momentum and energy conservation equations.

\subsection{Energy and flow equations in planar symmetric Lagrangian coordinates}

In Lagrangian coordinates, the matter continuity equation and the momentum conservation equation for $\rho$ and $v$, the density and velocity respectively, as a function of $x$, the position, and $t$, the time, are given by
\begin{align}
\frac{d\rho}{dt}=&-\rho \frac{dv}{dx}, \label{eq:continuity_equation} \\
\frac{dv}{dt}=&-\frac{1}{\rho} \frac{d}{dx} \left(p+\frac{1}{3}\int_0^\infty u_\varepsilon d\varepsilon\right), \label{eq:momentum_equation}
\end{align}
where $p$ is the matter pressure, $u_\varepsilon$ is the radiation energy density per unit photon energy (or spectral energy density), and the bolometric radiation pressure is proportional to bolometric energy density with a pre-factor of $1/3$.

The evolution in time of the spectral energy density and the matter energy density can be described schematically by
\begin{align}
\frac{du_\varepsilon}{dt}=&\frac{\partial u_\varepsilon}{\partial t}\bigg|_{\rm emit}+\frac{\partial u_\varepsilon}{\partial t}\bigg|_{\rm abs}+\frac{\partial u_\varepsilon}{\partial t}\bigg|_{\rm scat}+\frac{\partial u_\varepsilon}{\partial t}\bigg|_{\rm comp}+\frac{\partial u_\varepsilon}{\partial t}\bigg|_{\rm trans}, \label{eq:dudt} \\
\frac{de}{dt}=& -\int_0^\infty \left[\frac{\partial u_\varepsilon}{\partial t}\bigg|_{\rm emit}+\frac{\partial u_\varepsilon}{\partial t}\bigg|_{\rm abs}+\frac{\partial u_\varepsilon}{\partial t}\bigg|_{\rm scat}\right]d\varepsilon+\frac{\partial e}{\partial t}\bigg|_{\rm comp}, \label{eq:dedt}
\end{align}
where the terms on the right-hand side of eq.~\eqref{eq:dudt} describe emission, absorption, scattering, compression and radiation transport, respectively.
In this formulation, the radiation energy in different photon energies changes according to the radiative interaction with the matter (emission, absorption and scattering), as well as through transport and compression, and the change in the matter energy is the total energy transferred through the radiative interaction and the matter compression work.

For the non-relativistic regime considered, and considering an optically thick plasma, radiation transport can be described by the diffusion equation, and Compton scattering can be described by the Kompaneets equation \citep{Kompaneets57}.
In addition, as the radiation is isotropic in the matter rest frame, the radiation is directly coupled to the motion of matter.
This translates into a radiation compression work term, which in the comoving frame includes the Doppler and aberration corrections \citep{Castor07}.
The different energy transfer and work terms are therefore given by
\begin{equation}\label{eq:u_terms}
\begin{split}
\frac{\partial u_\varepsilon}{\partial t}\bigg|_{\rm emit}&= \rho \kappa_\varepsilon c B_\varepsilon(T),\\
\frac{\partial u_\varepsilon}{\partial t}\bigg|_{\rm abs}&= -\rho \kappa_\varepsilon c u_\varepsilon, \\
\frac{\partial u_\varepsilon}{\partial t}\bigg|_{\rm scat}&=\rho \kappa c \frac{\varepsilon}{m_{\rm e} c^2}\frac{\partial}{\partial \varepsilon}\left[T\frac{\partial}{\partial \varepsilon}(\varepsilon u_\varepsilon)+(\varepsilon-4T)u_\varepsilon+\frac{(hc)^3}{8\pi}\left(\frac{u_\varepsilon}{\varepsilon}\right)^2\right], \\
\frac{\partial u_\varepsilon}{\partial t}\bigg|_{\rm comp}&=-\left[\frac{4}{3}u_\varepsilon-\frac{1}{3}\frac{\partial (\varepsilon u_\varepsilon)}{\partial \varepsilon}\right]\frac{\partial v}{\partial x}, \\
\frac{\partial u_\varepsilon}{\partial t}\bigg|_{\rm trans}&= -\frac{\partial j_\varepsilon}{\partial x}, \\
\frac{\partial e}{\partial t}\bigg|_{\rm comp}&=-(p+e)\frac{\partial v}{\partial x},
\end{split}
\end{equation}
where $B_\varepsilon(T)$ is the Planck spectrum in terms of energy density per unit photon energy, given by
\begin{equation}\label{eq:Planck_spectrum}
B_\varepsilon(T)=\frac{8\pi}{(hc)^3}\frac{\varepsilon^3}{e^{\varepsilon/T}-1},
\end{equation}
$\kappa$ is the electron scattering opacity, $\kappa_\varepsilon$ is the (photon-energy dependent) absorption opacity and $j_\varepsilon$ is the spectral energy flux.

Assuming that the plasma is completely ionized, with $Z$ and $A$ being the atomic charge and atomic weight respectively, the plasma equation of state is that of an ideal gas, $p=(\gamma-1)e=(1+Z)\rho T/A m_{\rm p}$ where $e$ is the matter energy density, and $\gamma=5/3$ is the adiabatic index.
At the temperature range considered here bremsstrahlung dominates the absorption opacity, and
\begin{equation}
\begin{split}
\kappa_\varepsilon=&\left(\frac{8 \pi^3}{3}\right)^{1/2}\frac{Z^2}{A}\alpha_{\rm e}^{-2}\frac{\rho}{m_{\rm p}}r_{\rm e}^3\left(\frac{T}{m_{\rm e}c^2}\right)^{-1/2}\left(\frac{\varepsilon}{m_{\rm e}c^2}\right)^{-3} \\
&\cdot \left(1-e^{-\varepsilon/T}\right) g_{\rm ff}\kappa,
\end{split}
\end{equation}
where the electron scattering opacity is
\begin{equation}\label{eq:kappa_T}
\kappa=\frac{8\pi}{3}\frac{Z}{A}\frac{e^4}{m_{\rm p} (m_{\rm e} c^2)^2},
\end{equation}
and the Gaunt factor is given by \footnote{This expression approximates the asymptotic limits of the Gaunt factor in the Born approximation.} \citep{Rybicki86}
\begin{equation}
g_{\rm ff}=
\begin{cases}
\frac{\sqrt{3}}{\pi}\log\left(\frac{2.25 T}{\varepsilon}\right) &, \quad \varepsilon<T, \\
\frac{\sqrt{3}}{\pi}\log(2.25) \left(\frac{T}{\varepsilon}\right)^{1/2} &, \quad \varepsilon \geq T.
\end{cases}
\end{equation}
Using the diffusion approximation, the spectral energy flux is given by
\begin{equation}\label{eq:j_epsilon}
j_{\varepsilon}=-\frac{c}{3}\frac{\partial u_\varepsilon}{\partial \tau_\varepsilon},
\end{equation}
where the diffusion optical depth is $\tau_\varepsilon=\int_x \kappa_\varepsilon^* \rho dx$, and $\kappa_\varepsilon^*=\kappa+\kappa_\varepsilon$.

Note that for temperatures $<100 \eV$, the dominant absorption opacity at $\varepsilon>T$ is the recombination opacity and not bremsstrahlung.
This regime appears in the RMS problem as a precursor of high energy photons advances in a cold medium.
We discuss the validity of the opacity assumption in \sref{sec:Radiation mediated shock in a homogenous medium}.

\subsection{Numerical scheme for the solution of the equations}\label{sec:Numerical scheme for the solution of the equations}
The radiation-hydrodynamics equations, eqs.~\eqref{eq:continuity_equation}-\eqref{eq:dedt}, are solved in the following manner.
The continuity and momentum equations, eqs.~\eqref{eq:continuity_equation}-\eqref{eq:momentum_equation}, are solved by the standard leap-frog on a staggered-mesh.
The energy conservation set of equations is solved using operator splitting: the equations are divided into three parts, radiation transport, radiative processes, and radiation compression work, and these parts are solved consecutively.
Note that in the radiative processes part we include all the radiative interactions - emission, absorption and scattering.
The entire set of energy conservation equations is solved using a predictor-corrector in order to update the opacity values in the diffusion equation.

First, only the radiation transport term in eq.~\eqref{eq:dudt} is included in the radiation energy evolution.
The spectral energy density is advanced over a full time step, by solving the diffusion equation for each photon energy, using a photon energy dependent opacity and the appropriate boundary conditions.

Then, the spectral energy density solution coming out of the diffusion equation is fed into the radiative processes part, and advanced over a full time step.
In the radiative processes part, eq.~\eqref{eq:dudt} including only the emission, absorption and scattering terms, and the full eq.~\eqref{eq:dedt} are solved simultaneously, for each spatial cell.
For each time step this reduced set of equations is solved in an iterative fashion, as follows: the radiative processes part of eq.~\eqref{eq:dudt} is solved implicitly using the "modified Youngs" method reported in \citet{Larsen85}, for a given matter temperature at the end of the time step.
This method ensures a non-negative spectral energy density, without limiting the time step.
Then, the corresponding radiative energy transfer is introduced to the matter energy equation, eq.~\eqref{eq:dedt}, which includes the matter compression work.
The solution for the matter temperature is then inserted again into the reduced radiation energy equation, until convergence is achieved, and the spectral radiation energy density and the matter energy density solve simultaneously the reduced set of equations.
In this way, energy conservation in the radiative interactions between the radiation and the matter is assured to numerical precision.

Then, the spectral energy density solution coming out of the radiative processes part is fed into the radiation compression part, and advanced over a full time step.
In the radiation compression work part, only the compression work term in eq.~\eqref{eq:dudt} is included in the radiation energy evolution, and solved for each spatial cell.
The numerical procedure we employ to solve this equation is to reduce the spectral equation to the following form,
\begin{equation}\label{eq:rad_compress}
\left.\frac{\partial u_\varepsilon}{\partial t}\right|_{\rm comp}=-u_\varepsilon \left(1-\frac{1}{3}\frac{\partial \log u_\varepsilon}{\partial \log \varepsilon}\right) \frac{\partial v}{\partial x},
\end{equation}
and evaluate the logarithmic derivative explicitly by a high order central difference first derivative.
The other terms in the equation are evaluated and solved implicitly.

We note that the usual method of dealing with the compression part in eq.~\eqref{eq:u_terms} is by using a first order forward (or backward) difference first derivative to evaluate the spectral partial derivative term.
In this way, a telescopic sum over the spectral partial derivative term cancels its bolometric contribution, and total energy is conserved by definition.
Although the method used in this work has a disadvantage over the usual method as energy conservation has a truncation error determined by the time step and the spectral grid spacing, it achieves better accuracy in the spectral energy density while requiring much less resolution in spectral points.
This property is particularly important in describing the exponential tail of the spectral energy.
For instance, in adiabatic expansion the radiation keeps a Planckian shape, and a method that uses a first order finite difference first derivative fails to describe this without resorting to great spectral resolution.

After the radiation compression part is solved and the spectral energy density is provided, the entire energy set of equations is solved again for the same time step, with the opacity values in the diffusion equation updated according to the change in the matter temperature.
Following convergence, the time step is advanced, starting again from the momentum and continuity equations.

\subsection{The simulation time step}\label{sec:The simulation time step}

At the end of each time step the following time step is evaluated according to the hydrodynamic and radiation conditions.
The time step is determined as the minimum between $\Delta t_{\rm c}$, the Courant time step, $\Delta t_{\rm d}$, the diffusion time step and $\Delta t_{\rm r}$, the radiative processes time step.
The Courant time step is the usual minimum of $\Delta t_{\rm c}=f_{\rm c} \Delta x/C_{\rm s}$ over the spatial cells where $\Delta x$ is the grid spacing, $C_{\rm s}$ is the total local sound speed and $f_{\rm c}$ is some positive fraction $<1/3$.
The diffusion time step is determined by the energy flux in the boundary cells - note that an implicit solution of the diffusion equation does not ensure non-negative energies if an energy flux boundary condition is imposed.
This time step is calculated as the minimum (over photon energies and over the boundary cells where $\partial j_\varepsilon/\partial x>0$) of
\begin{equation}
\Delta t_{\rm d}=f_t \frac{u_\varepsilon}{\partial j_\varepsilon/\partial x},
\end{equation}
where $f_t$ is some positive fraction $<0.1$.
The radiative processes time step is chosen to ensure non-negative matter and radiation energies, and it is the minimum (over grid cells) of
\begin{equation}
\Delta t_{\rm r}=f_t
\begin{cases}
\frac{u}{j_{\rm B}+j_{\rm C}} &, \quad j_{\rm B}+j_{\rm C}<0, \\
\frac{e}{j_{\rm B}+j_{\rm C}} &, \quad j_{\rm B}+j_{\rm C}>0,
\end{cases}
\end{equation}
where $u=\int_0^\infty u_\varepsilon d\varepsilon$ is the bolometric radiation energy density, the effective plasma emissivity is given by
\begin{equation}\label{eq:emissivity_B}
j_{\rm B}=\rho c \int_0^\infty \kappa_\varepsilon\left(B_\varepsilon-u_\varepsilon\right)d\varepsilon,
\end{equation}
and the Compton scattering emissivity of the plasma can be estimated as
\begin{equation}\label{eq:emissivity_C}
j_{\rm C}=4 u (\rho \kappa c)  \frac{T-T_\gamma}{m_{\rm e} c^2},
\end{equation}
where the radiation temperature is defined as \citep{Zel'dovich70}
\begin{equation}\label{eq:T_gamma}
T_{\gamma}=\frac{1}{4u}{\int_0^\infty \left[\varepsilon u_\varepsilon+\frac{(hc)^3}{8\pi}\left(\frac{u_\varepsilon}{\varepsilon}\right)^2\right]d\varepsilon}.
\end{equation}
This definition of the radiation temperature holds the important property of the radiation and the matter having the same temperature both at LTE and at Compton equilibrium.
Moreover, this definition preserves the correct scaling, $T_\gamma \propto \rho^{1/3}$, in adiabatic expansion/compression, even for a non-Planckian energy spectrum (see \sref{sec:Properties of the radiation temperature}).

A numerical code implementing this scheme was written and verified against simple benchmark problems that have bolometric and spectral analytical solutions (see \sref{sec:Test problem for the numerical code}).
An additional problem that has an analytical bolometric solution is the steady-state RMS.
This problem is reviewed next, before addressing the problem of a planar RMS breaking out from a stellar surface.

\section{Radiation mediated shocks in a homogenous medium}\label{sec:Radiation mediated shock in a homogenous medium}

The problem of a steady RMS propagating in a homogenous medium was extensively studied in the literature \citep[e.g.][]{Weaver76,Blandford81,Lyubarskii82,Becker88,Riffert88}.
Analytical solutions for the hydrodynamic and bolometric properties of a strong RMS were provided in \citet{Weaver76}, as well as approximate solutions for the temperature profile using the effective photon approximation.
In this approximation the radiation and the matter are assumed to be in local Compton equilibrium, and the photon production rate includes all photons that can saturate in energy by Comptonization.
Here we test the spectral calculations performed with the scheme described in the previous section vs. the analytical hydrodynamic solution and the temperature solution in the effective photon approximation.

\subsection{Initial conditions and boundary conditions}\label{sec:Initial conditions and boundary conditions}
The calculation is performed with a constant velocity boundary condition and a uniform density matter, which simulates a piston driven into a homogeneous medium.
The parameters that determine the problem are $v_0$, the shock velocity, $\rho_0$, the pre-shock density and the matter composition.
In order to facilitate fast convergence, the initial conditions are taken from the analytical hydrodynamic profile, and the radiation and the matter are assumed to be in LTE.

Denoting the shock's position in terms of the Lagrangian mass coordinate (per unit area) as $m_{\rm sh}$, the initial density, velocity and radiation pressure as a function of mass are taken to be $\rho_{\rm W}(m-m_{\rm sh})$, $v_{\rm W}(m-m_{\rm sh})$ and $p_{\rm W}(m-m_{\rm sh})$, defined in eq.~\eqref{eq:RMS_structure} in \sref{sec:Structure of a radiation mediated shock}.
The radiation (and matter) temperature is taken as $T(m)=[3p_{\rm W}(m-m_{\rm sh})/a_{\rm BB}]^{1/4}$, and the radiation spectrum is taken as a black-body spectrum, $u_\varepsilon(m)=B_\varepsilon(T(m))$, where $a_{\rm BB}=8\pi^5/15(hc)^{3}$.
The velocity at the piston boundary is taken as $(6/7)v_0$, and following the constant velocity condition, a null energy flux is imposed.
A reflective boundary condition is taken at the outer boundary, without affecting the calculation.

Reported here are calculations with a box size of $100 m_0$ and a resolution of $m_0/20$, where $m_0=(\kappa \beta_0)^{-1}$, and the shock's initial position is taken to be at $m_{\rm sh}=10 m_0$, and moving in the direction of positive mass.
The spectral grid is taken between photon energies of $1 \meV$ and $10^3 [(18/7)\rho_0 v_0^2/a_{\rm BB}]^{1/4}$, and divided logarithmically to $200$ spectral points.
A minimal temperature of $\eV/100$ was imposed in the far upstream for numerical purposes, with negligible effect on the results.

After the shock has propagated $\approx 50 m_0$ in mass, the hydrodynamic and temperature profiles stop developing with time, in the shock's frame, and a steady-state is obtained.
These spatial and spectral resolutions are converged to an error $<1\%$ in the peak temperature, as well as in the temperature of the radiation precursor, described next.

\subsection{Temperature results}\label{sec:Temperature results}
\begin{figure}[h]
\includegraphics[scale=0.8]{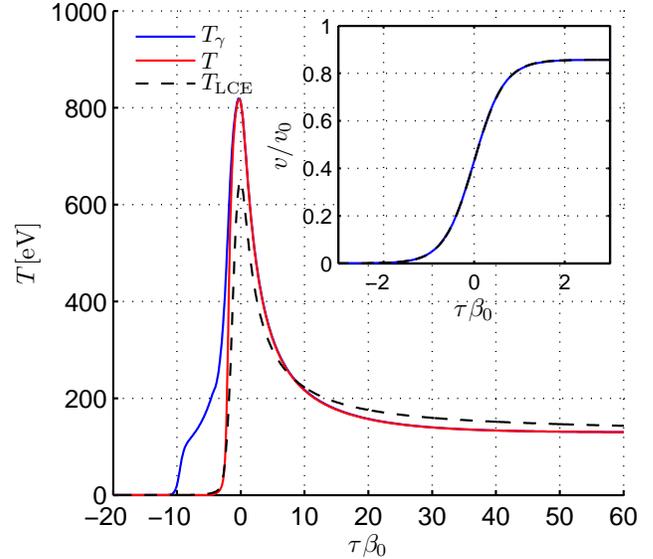}
\caption{\label{fig:T_vs_tau_RMS}
Temperature profile of a steady-state RMS as a function of optical depth around the shock's position. The presented profile is for a calculation with a normalized shock velocity of $\beta_0=v_0/c=0.1$ in a hydrogen medium with a pre-shock proton number density of $n_{\rm p,0}=10^{15} \cm^{-3}$.
The zero optical depth is set to the point where the bolometric energy flux is at its maximum, and the optical depth is normalized by $\beta_0^{-1}$. The shock is shown to be moving in the direction of negative optical depth. Presented are the radiation temperature profile according to eq.~\eqref{eq:T_gamma} (solid blue) and the matter temperature (solid red). Also presented is the temperature profile calculated in the effective photon approximation (dashed black).
The inset shows the (absolute) velocity profile as a function of optical depth. Presented are the calculated profile (dashed blue) and the analytical profile according to eq.~\eqref{eq:RMS_structure} (solid black).
}
\end{figure}

Figure \ref{fig:T_vs_tau_RMS} presents the radiation temperature and the matter temperature profiles as a function of optical depth, in a spectral calculation with $\beta_0=0.1$ and a proton number density of $n_{\rm p,0}=\rho_0/m_{\rm p}=10^{15}\cm^{-3}$ in a hydrogen plasma (calculations for any hydrogen-helium mixture provide the same results).
The figure's inset presents the calculated and the analytical velocity profiles, showing excellent agreement ($<10^{-3}\%$ error).
The zero position in the plots is set to the position of maximum energy flux.
As can be seen, the shock has a high temperature radiation precursor, extending to an optical depth of $\approx 10 \beta_0^{-1}$ in advance of the shock's position.
This precursor carries negligible energy, but the ratio of the photon number density to the ion number density is $n_\gamma/n_{\rm p,0}>1$ up to at least an optical depth of $\approx 4 \beta_0^{-1}$ ahead of the shock, where $n_\gamma=\int_0^\infty u_\varepsilon/\varepsilon d\varepsilon$.
Thus there are enough photons to ionize the material ahead of the shock, but beyond an optical depth of $\approx 4 \beta_0^{-1}$ into the upstream the precursor is not described correctly.

Note that the radiation precursor propagates in a cold plasma, with temperature $<100 \eV$.
For this temperature range the recombination opacity dominates over bremsstrahlung for photons with $\varepsilon>T$, and was not included in the calculations appearing here.
However, the recombination absorption opacity is negligible compared to $\kappa$ for $T>10 \eV$ and $n_{\rm p,0} \leq 10^{15} \cm^{-3}$, which are the conditions at an optical depth of $\approx 4\beta_0^{-1}$ from the shock's front, and the effective optical depth for recombination absorption is negligible with respect to $\beta_0^{-1}$.
Therefore, at optical depths $\lesssim 4\beta_0^{-1}$ from the shock front the precursor is not expected to be be significantly modified by recombination absorption.
We confirmed this with calculations including the recombination opacity.

Near the position of maximum energy flux the radiation temperature reaches a peak value, and the matter and the radiation roughly equilibrate their temperatures.
The exact point where the radiation temperature peaks is where the velocity divergence peaks, at an optical depth of $0.32\beta_0^{-1}$ upstream to the zero position.
That is not a mere coincidence, and we elaborate on this point below.
Further downstream, the two temperatures are coupled together and decrease as the radiation and matter slowly approach LTE.
Also presented in the main figure is the temperature profile obtained in the effective photon approximation.
As can be seen, the discrepancy between the approximate and the exact solutions is $21\%$ at peak temperature.
This discrepancy increases for higher shock velocities, but even for $\beta_0=0.2$ it is still only $27\%$.
For $\beta_0=0.05$ the difference in the peak temperature between the calculations is $<3\%$.

\begin{figure}[h]
\includegraphics[scale=0.8]{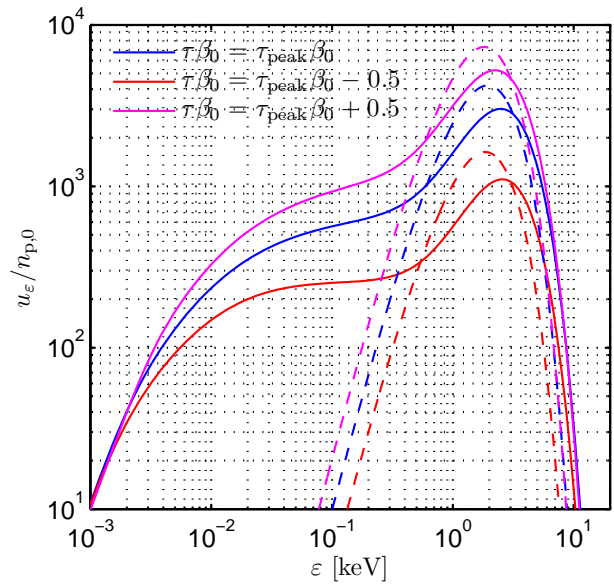}
\caption{\label{fig:RMS_spectrum}
Spectral energy density as a function of photon energy, in the vicinity of the RMS's position. The spectral energy density is normalized to $n_{\rm p,0}$, the pre-shock proton number density. The different lines present the spectrum at the zero position (solid blue), at an optical depth $0.5\beta_0^{-1}$ upstream to the zero position (solid red) and $0.5\beta_0^{-1}$ downstream to the zero position (solid magenta).
Also presented are Wien spectra calculated in the effective photon approximation at the same positions (dashed lines, same color scheme).
}
\end{figure}

Figure \ref{fig:RMS_spectrum} presents the spectral energy density as a function of photon energy in the vicinity of the shock front: at the zero position, and at optical depths of $-0.5\beta_0^{-1}$ and $+0.5\beta_0^{-1}$ around the zero position.
As can be seen, the spectra around the peak flux position resemble Comptonized spectra \cite[e.g.][]{Felten72,Illarionov72}.
Also shown in the plot are Wien spectra, from calculations with the effective photon approximation.
While the area under the curve (the bolometric energy density) is similar between the different calculations for each position, the peak temperature and the photon number density are different.
This can be easily seen as the spectral energy density is equal to the photon number density per logarithmic unit of photon energy.

Notably, the radiation spectrum does not feature a power-law tail produced by bulk Comptonization, as previously proposed \citep{Blandford81,Suzuki10}.
In this process photons gain energy in electron scattering due to the relative bulk motion of electrons in the shock front (this process is described in the equations by the Doppler and abberation corrections in the compression term in eq.~\eqref{eq:u_terms}).
To gain a rough estimate for this effect, let us assume a velocity jump of $\Delta v=(6/7)\beta_0 c$ spread over an optical depth of $\tau=3\beta_0^{-1}$.
The energy gain is therefore $\sim 0.3 \beta_0^2 \varepsilon$ per scattering (this is a converging flow - scattered photons are blue-shifted, whether they come from the upstream or the downstream).
Photons also gain (lose) energy due to the electrons' thermal motion (Compton recoil), with a net energy gain of $\sim (4T-\varepsilon)\varepsilon/m_{\rm e}c^2$ per scattering.
For an isothermal shock, these processes are balanced for photons with energy $\varepsilon \sim 4T+0.3 \beta_0^2 m_{\rm e}c^2$, and there can be no energy gain above this cutoff energy.
Thus, at energies above this cutoff the spectrum falls exponentially, but below this energy the spectrum is described by a power-law down to the downstream temperature \citep{Blandford81,Lyubarskii82}.

So why does a Comptonized spectrum actually form?
Photons diffusing from the downstream can (adiabatically) gain energy only where there is a velocity divergence.
Since the radiation is compressed at the shock front, the declining branch of the spectrum ($d(\varepsilon u_\varepsilon)/d\varepsilon<0$) increases in energy at the expense of the inclining branch.
This already enhanced radiation is advected with the flow to the downstream of the shock, diffusing over the shock's front and gaining more energy.
Photons as far as $7\beta_0^{-1}$ downstream to the shock front can reach the shock by diffusion, and the process of photon "acceleration" then repeats itself.
If electron scattering was purely elastic in the matter rest frame and there was no other energy exchange between the radiation and the matter, the spectrum solution in steady-state would be a (declining) power-law in energy \citep{Blandford81} \footnote{Obviously, this is not a self-consistent picture.
Without coupling between the radiation and the matter nothing constrains the matter temperature, and a seed radiation spectrum needs to be assumed for the downstream.
Also, if photons are not continuously produced at the downstream, the RMS will fade away and turn into a ion-viscous shock.}.
However, at the downstream, where there is no compression, photons with $\varepsilon>4T$ ($\varepsilon<4T$) simply lose (gain) energy in electron scattering, and Compton equilibrium is established between the radiation and the matter with the spectrum redistributing to the familiar Wien spectrum \citep[see also discussions in][]{Blandford81,Becker88,Riffert88}.

Nonetheless, the bulk Comptonization process is important for determining the temperature profile.
If the compression term did not include the Doppler and aberration corrections, the radiation temperature could not exceed the downstream temperature anywhere, and no temperature peak would form.
Essentially, the two processes, bulk and thermal Comptonization, are competing where there is significant compression.
But the compression, determined by the velocity divergence, is not uniform on the shock front.
At the point of highest compression, where the velocity divergence is at its maximum, the peak temperature is obtained.
Further downstream, thermal Comptonization is stronger and the temperature declines.

In conclusion, we have shown that the calculated hydrodynamic profiles agree with the analytical results.
In fact, he basic equation solved by \cite{Weaver76} is obtained by combining the energy equations (eqs.~\eqref{eq:dudt} and \eqref{eq:dedt}) and integrating over photon energy, considering that electron scattering dominates the opacity and that the matter pressure is negligible.
Thus, any single or multi-group treatment that includes diffusion and compression should yield the correct bolometric result.

We have also shown that the effective photon approximation describes well the radiation temperature (up to an error of $\sim 30\%$).
The success of the effective photon approximation relies on three properties.
First, the radiation and the matter are assumed to be at the same temperature.
Second, the radiation is assumed to be described by a Comptonized spectrum, with a Wien component at the peak.
Third, the effective photon production rate is estimated locally and includes all the photons that can be up-scattered till saturation in energy.
The first two properties relate the matter temperature, the radiation energy density and the photon number density through the ratio $u=3n_\gamma T$, which is assumed to hold everywhere.
The third property replaces the photon absorption and emission terms with an effective photon production rate.

In fact, the equation for the photon number density in the effective photon approximation is the same as the one in the spectral calculations, only with an effective photon production rate in place of the absorption and emission terms.
Hence, if all these assumptions hold, the effective photon approximation should yield exactly the same temperature as in the spectral calculations.
The main reason this approximation does not produce the exact same temperature as in a detailed calculation is the inaccuracy of the effective production rate.
As photons scatter and diffuse in space before saturating in energy, the locally estimated cutoff energy in the effective production rate is not accurate.
In addition, the peak of the radiation spectrum becomes wider at lower densities, and is not described well by a Wien Component.
This leads to an inexact photon number density and an inexact radiation temperature.

\section{Planar shock breakout from a stellar surface}\label{sec:Planar shock breakout from a stellar surface}

The radiation emission at the RMS breakout depends on the shock dynamics in the envelope and on the shock's structure.
In a hydrostatic stellar envelope, the density profile near the surface can be described as a power-law of the distance from the surface.
A shock propagating outwards in such a density profile accelerates and is not sensitive to its initial driver, approaching a self-similar solution \citep{Gandel'Man56,Sakurai60}.
On the other hand, the width of the shock depends on the pre-shocked density, the shock velocity and the opacity.
Thus, the problem of shock breakout from a stellar surface in the planar approximation is defined by the parameters describing the density profile and the asymptotic shock velocity, $\rho_0$ and $\beta_0$, and the matter composition, determining the electron scattering opacity.
Ignoring any circumstellar or interstellar material, the density profile prior to the shock's passage can be defined by
\begin{equation}\label{eq:density_profile}
\rho_{\rm i}(\tau)=\rho_0 (\beta_0 \tau)^{n/(n+1)},
\end{equation}
where $\tau$ is the optical depth of the matter lying between the stellar surface and the layer position $x$,
\begin{equation}
\tau=\int_{0}^{x}\kappa\rho dx=\kappa m,
\end{equation}
where $m$ is the mass per unit area and $\kappa$ is the electron scattering opacity, defined in eq.~\eqref{eq:kappa_T}.
At large optical depth, the shock velocity as a function of the optical depth ahead of the shock front is described by
\begin{equation}\label{eq:shock_velocity_acc}
v_{\rm acc}(\tau)=-v_0 (\beta_0 \tau)^{-\lambda/(n+1)},
\end{equation}
where $\lambda\approx 0.19 n$ is found with a self-similar analysis (see \sref{sec:Hydrodynamic profiles in the self-similar solution}).

We remind here the scaling of the bolometric properties of the burst with the breakout parameters.
The total energy emitted in the prompt burst is roughly the internal energy deposited in the shock's passage in the emitting layer, which has a size comparable to the width of the shock's front \citep{Matzner99,Katz10,Nakar10,Piro10,Sapir11,Katz12}.
For a star with a radius $R$, the mass contained in an optical depth $\tau$ is $M \sim 4\pi R^2 \tau/\kappa$, and the total emitted energy is estimated as $\sim R^2 \tau v^2/\kappa$.
The characteristic fluence (emitted energy per unit area) is therefore
\begin{equation}\label{eq:e0}
\mathcal{E}_0=\frac{\beta_0 c^2}{\kappa}.
\end{equation}
Locally, at each point on the surface, the energy flux scaling is determined by the kinetic energy flux in the shock,
\begin{equation}\label{eq:L0}
\mathcal{L}_0=\rho_0 (\beta_0 c)^3,
\end{equation}
and the emission time scale is similar to the shock crossing time,
\begin{equation}\label{eq:t0}
t_0=\frac{c}{\rho_0 \kappa \beta_0^2}.
\end{equation}
These values are not to be confused with the observed flux and observed emission time scale, which are determined by the light travel time $R/c$
\citep[An exact calculation of the bolometric fluence and flux emitted in the burst is reported in][]{Sapir11,Katz12}.

In order to achieve fast convergence in the spectral calculations, the inner boundary conditions are determined by the pure hydrodynamic self-similar solution, and the initial conditions are taken from an ansatz combining the pure hydrodynamic solution of a discontinuous shock with the structure of a steady RMS \citep[see discussion in][]{Sapir11}.

\subsection{Initial conditions}\label{sec:Initial conditions}
The time when the shock emerges at the surface in the pure hydrodynamic solution is defined here as $t=0$.
At $t_{\rm init}<0$, the initial time of the calculation, the shock's position is taken to be at an optical depth $\tau_{\rm sh}=\kappa m_{\rm sh}$.
The initial hydrodynamical conditions are taken from an ansatz combining the solutions of the self-similar accelerating shock and the steady-state RMS, assuming negligible matter pressure.
Taking $\rho_{\rm sh}=\rho_{\rm i}(m_{\rm sh})$ and  $v_{\rm sh}=v_{\rm acc}(m_{\rm sh})$ to be the pre-shock density and the shock velocity at the shock's position, respectively, the velocity, radiation pressure and spatial position, as a function of mass from the surface, are given by
\begin{equation}\label{eq:breakout_initial_conditions}
\begin{split}
v(m,t_{\rm init})=&
\begin{cases}
v_{\rm sh}\frac{v_{\rm W}(m_{\rm sh}-m)}{v_0} &, \quad m \leq m_{\rm sh}, \\
v_{\rm S}(m,t_{\rm init})\frac{v_{\rm W}(m_{\rm sh}-m)}{v_0} &, \quad m>m_{\rm sh}
\end{cases}, \\
p_r(m,t_{\rm init})=&
\begin{cases}
\rho_{\rm sh}v_{\rm sh}^2\frac{p_{\rm W}(m_{\rm sh}-m)}{\rho_0 v_0^2} &, \quad m \leq m_{\rm sh}, \\
p_{\rm S}(m,t_{\rm init})\frac{p_{\rm W}(m_{\rm sh}-m)}{\rho_0 v_0^2} &, \quad m>m_{\rm sh}
\end{cases}, \\
x(m,t_{\rm init})=&x_{\rm S}(m,t_{\rm init})+x_{\rm W}(m_{\rm sh}-m) \\
&-
\begin{cases}
x_{\rm W,m \rightarrow \infty}(m) &, \quad m \leq m_{\rm sh}, \\
x_{\rm W,m \rightarrow -\infty}(m) &, \quad m>m_{\rm sh}
\end{cases},
\end{split}
\end{equation}
where the self-similar and the steady-state RMS functions are provided in \sref{sec:Structure of a radiation mediated shock} and \sref{sec:Hydrodynamic profiles in the self-similar solution}.
The initial matter density is calculated according to $\rho(m,t_{\rm init})=\partial m/ \partial x(m,t_{\rm init})$, and the initial temperature profile is taken from the black-body value of $T(m,t_{\rm init})=[3p_r(m,t_{\rm init})/a_{\rm BB}]^{1/4}$. Accordingly, the radiation is started with a black-body spectrum, $u_\varepsilon(m,t_{\rm init})=B_\varepsilon(T(m,t_{\rm init}))$, and the matter pressure is taken as $p(m,t_{\rm init})=(1+Z)\rho(m,t_{\rm init})T(m,t_{\rm init})/A m_{\rm p}$.
The minimal temperature in the far upstream is taken as $0.1 \eV$, but this has little effect on the results.
Note that for high values of density and low values of shock velocity where the matter pressure is not negligible, this ansatz is not self-consistent.

\subsection{Boundary conditions}\label{sec:Boundary conditions}
The inner boundary in the calculations, at an optical depth $\tau_{\rm R}$, represents the far downstream of the accelerating shock where diffusion is negligible in comparison to adiabatic expansion.
Accordingly, the conditions at the inner boundary can be determined from the self-similar solution, which is described in \sref{sec:Hydrodynamic profiles in the self-similar solution}.
The time dependent velocity of the inner boundary, $v_{\rm S}(\tau_{\rm R},t)$, is taken from this solution.

In the far downstream the matter and the radiation are in LTE, and the radiation is characterized by a black-body spectrum.
In addition, the far downstream is optically thick, so the diffusion approximation is appropriate.
Thus, for a given bolometric flux the spectral flux at the inner boundary can be determined by the LTE condition together with the diffusion approximation.

Following the definition of the spectral flux in the diffusion approximation, eq.~\eqref{eq:j_epsilon}, the spectral flux can be separated to the different partial derivatives using the chain rule,
\begin{equation}\label{eq:inner_boundary_spectral_flux}
j_{\varepsilon}=-\frac{c}{3}\frac{\partial u_\varepsilon}{\partial T}\frac{\partial T}{\partial u}\frac{\partial u}{\partial \tau}\frac{\partial \tau}{\partial \tau_\varepsilon}.
\end{equation}
Assuming a black-body spectrum for the radiation, and assuming that electron scattering dominates the opacity, the following relations apply
\begin{equation}\label{eq:Par_deriv}
\begin{split}
\frac{\partial u}{\partial \tau}&=-\frac{3}{c}j, \\
\frac{\partial u_\varepsilon}{\partial T}&= B_\varepsilon(T)\frac{\varepsilon}{T^2}\frac{e^{\varepsilon/T}}{e^{\varepsilon/T}-1}, \\
\frac{\partial u}{\partial T} &= \frac{32\pi^5 T^4}{15(hc)^3}, \\
\frac{\partial \tau_\varepsilon}{\partial \tau} &= \frac{\kappa_\varepsilon^*(\rho,T)}{\kappa},
\end{split}
\end{equation}
where $j=\int_0^\infty j_\varepsilon d\varepsilon$ is the bolometric flux.
To calculate the spectral flux at the inner boundary according to eq.~\eqref{eq:inner_boundary_spectral_flux}, the density and the temperature in eq.~\eqref{eq:Par_deriv} are evaluated at the boundary from the (time-dependent) pure hydrodynamic self-similar solution, where it is assumed that the energy density is dominated by radiation in LTE and that the matter pressure is negligible.
In addition, the bolometric energy flux at the boundary is calculated from the acceleration in the self-similar solution,
\begin{equation}
j_{\rm R}(t)=\frac{c}{\kappa}a_{\rm S}(\tau_{\rm R},t).
\end{equation}
For the outer boundary condition, defined at the surface at $\tau=0$ where the diffusion approximation does not apply, an Eddington relation is assumed
\begin{equation}\label{eq:j_outer}
j_{\varepsilon,\rm L}(t)=f_{\rm edd} u_\varepsilon (0,t) c,
\end{equation}
where $f_{\rm edd}$, the Eddington factor, depends on the angular distribution of the radiation intensity at the surface, and is assumed here to be photon energy independent.
While a full radiation transport calculation is needed in order to evaluate this parameter, the sensitivity to this factor determines whether the diffusion approximation is justified.
As the flux at the boundary is produced at an optical depth of $\tau \approx \beta_0^{-1}$ and is $j \approx \rho v_0^3$, the energy density at the boundary is lower than the energy density inside by a factor $f_{\rm edd}^{-1}\beta_0$.
This immediately implies that the luminosity sensitivity to the Eddington factor would be low in calculations with non-relativistic velocities.
However, the assumption of a photon energy independent Eddington factor implies that the calculated emitted flux is not exact at energies where absorption dominates over scattering.

Considering that photons lose all their momentum in absorption and (statistically) in forward-backward symmetric scattering, the velocity at the outer boundary is determined by the radiation momentum deposited in the ions,
\begin{equation}
\frac{dv_{\rm L}}{dt}=\int_0^\infty \frac{\kappa_\varepsilon^*}{c}j_{\varepsilon,\rm L}d\varepsilon-\frac{dp}{dm}(0,t).
\end{equation}
Note that this form is also correct where the diffusion approximation applies.
Also, there are assumed to be zero mass and matter pressure outside the surface.

\subsection{Temperature profiles and spectral luminosity}\label{sec:Temperature profiles and spectral luminosity}
Reported here are calculations with a non-uniform grid, where the $(i+1)$th cell's mass is taken to be $\Delta m_{i+1}=m_0(\Sigma_1^{i+1} \Delta m_{i}/m_0)^{\lambda/(n+1)}/N_{\rm sh}$, where $m_0=(\kappa \beta_0)^{-1}$ and the initial cell mass is $\Delta m_{1}=m_0/30$.
Such a grid ensures that the shock's front is resolved with $N_{\rm sh}$ cells throughout the shock's propagation.
The total mass for $N$ spatial cells is roughly (for large values of $N$)
\begin{equation}
M=\left(\frac{1+n-\lambda}{n+1} \frac{N}{N_{\rm sh}} \right)^{(n+1)/(1+n-\lambda)}m_0,
\end{equation}
and the shock's initial position is taken to be at $m_{\rm sh}=M/2$.
This allows a large enough optical depth in the downstream so the assumption of LTE at the boundary is valid.
The spatial grid was taken with $N=800$ cells and $N_{\rm sh}=20$, and the spectral grid was divided to $200$ spectral points, distributed logarithmically between photon energies of $0.01 \eV$ and $10^3 [(18/7)\rho_0 v_0^2/a_{\rm BB}]^{1/4}$.
These spatial and spectral resolutions are converged to an error $<1\%$ in the peak temperature.
Next are presented calculations for a hydrogen envelope with a density index $n=3$.
An Eddington factor of $f_{\rm edd}=1$ (see eq.~\eqref{eq:j_outer}) was taken in all the calculations \citep[for comparison, for a Thomson scattering envelope in the constant flux limit $f_{\rm edd}=0.71$, see][]{Katz12}.

An important result of the breakout model is that the spectral fluence, which is the emitted spectral flux integrated over time,
\begin{equation}
\mathcal{E}_\varepsilon=\int_{-\infty}^\infty j_{\varepsilon,\rm L}(t) dt,
\end{equation}
is a robust estimate not sensitive to the density slope, to the angular distribution of the radiation intensity, to the light travel time details or to mild asphericity of the explosion \citep{Sapir13}.
In the effective photon approximation, assuming Wien shape for the spectral flux, the spectral fluence obtains a universal shape, not sensitive to the values of $\beta_0$ and $\rho_0$.
We review next some of the results from the spectral calculations and compare them to the ones obtained in the effective photon approximation, with emphasis on the variables that determine the spectral fluence.

\begin{figure}[h]
\includegraphics[scale=0.8]{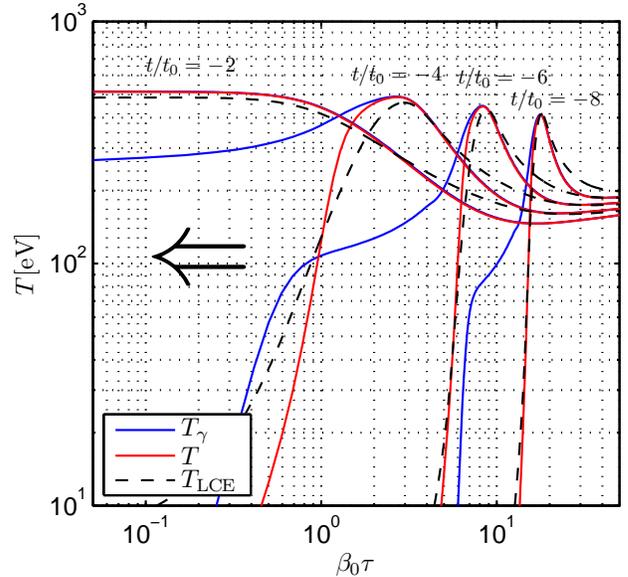}
\caption{\label{fig:T_vs_tau}
Temperature profiles in an RMS breakout as a function of optical depth from the surface at different times (shown in text over the curves).
Presented are the radiation temperature according to eq.~\eqref{eq:T_gamma} (solid blue), the matter temperature (solid red) and the temperature profile in the effective photon approximation \citep{Sapir13} (dashed black), calculated for a hydrogen envelope with shock velocity $\beta_0=0.1$ and proton number density $n_{\rm p,0}=10^{15} \cm^{-3}$.
The black arrow shows the shock's direction.
}
\end{figure}

Figure \ref{fig:T_vs_tau} presents the radiation and matter temperature profiles at different times prior to breakout in a calculation with $\beta_0=0.1$ and $n_{\rm p,0}=\rho_0/m_{\rm p}=10^{15} \cm^{-3}$.
As can be seen, similar to the steady-state case, there is a radiation precursor ahead of the shock's front, again carrying little energy.
Behind the point where the radiation reaches peak temperature the matter and radiation have equalized temperatures.
Following breakout, $t\gtrsim -2t_0$, the radiation and the matter have the same temperature everywhere.
Also presented is the temperature calculated in the effective photon approximation, showing very good agreement with the matter temperature.

\begin{figure}[h]
\includegraphics[scale=0.8]{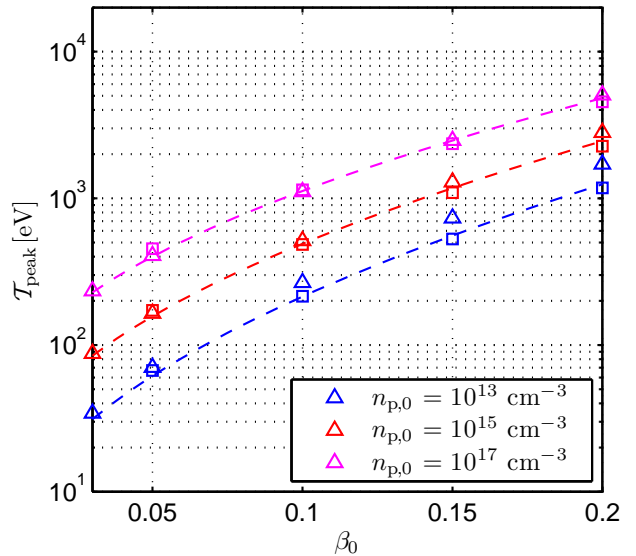}
\caption{\label{fig:Tpeak_vs_bt0_np0_n3}
Peak surface temperature as a function of breakout shock velocity $\beta_0$ for different values of pre-breakout proton number density $n_{\rm p,0}$ in a hydrogen envelope.
The results of the spectral calculations for $n=3$ are presented by triangles, the results of the photon effective approximation are presented by squares, and the fitting formula from \cite{Sapir13} is shown in dashed lines.
}
\end{figure}

Figure \ref{fig:Tpeak_vs_bt0_np0_n3} presents $\mathcal{T}_{\rm peak}$, the peak radiation temperature at the surface, in calculations with different values of $\beta_0$ and $n_{\rm p,0}$, in a hydrogen envelope (calculations for any hydrogen-helium mixture provide the same results).
For comparison, also presented are the peak temperatures from calculations with the effective photon approximation and the fitting formula reported in \cite{Sapir13}.
As can be seen, the effective photon approximation agrees with the exact spectral calculations, with great accuracy, at low shock velocities.
As the shock velocity is higher, and the density is lower, so is the discrepancy larger between the different calculations.
This behavior is similar to the one obtained in the case of the steady-state RMS.
But even for breakout parameters representing the upper limit of blue supergiants, for which high shock velocities can be achieved at low densities, e.g. $\beta_0=0.2$ and $n_{\rm p,0}=10^{13} \cm^{-3}$ \citep[see][]{Katz12}, the difference in peak temperature between the spectral calculations and the effective photon approximation is only $30\%$.

\begin{figure}[h]
\includegraphics[scale=0.8]{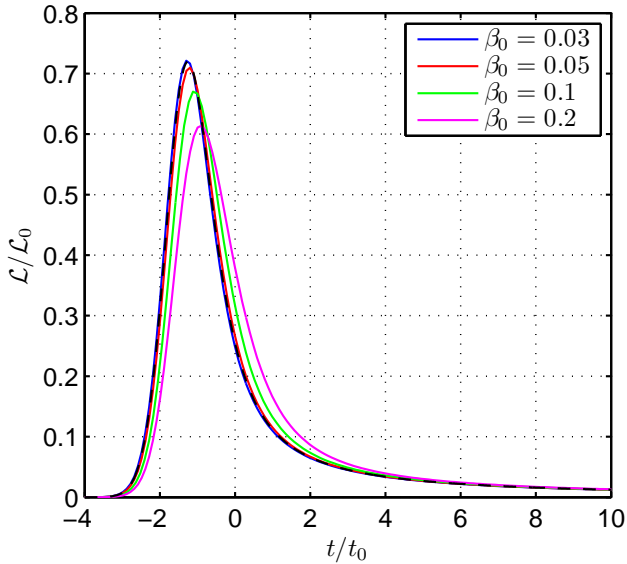}
\caption{\label{fig:Bol_luminosity}
Emitted energy flux as a function of time for different values of $\beta_0$. The flux is normalized by $\mathcal{L}_0$ (eq.~\eqref{eq:L0}) and the time is normalized by $t_0$ (eq.~\eqref{eq:t0}).
Solid lines present spectral calculations for a hydrogen envelope with different values of $\beta_0$, and $n_{\rm p,0}=10^{15} \cm^{-3}$, assuming free streaming at the surface ($f_{\rm edd}=1$).
The dashed black line present the free surface limit of $u(\tau=0)=0$.
}
\end{figure}

Figure \ref{fig:Bol_luminosity} presents $\mathcal{L}=\int_{0}^\infty j_{\varepsilon,\rm L}(t) d\varepsilon$, the bolometric energy flux at the surface, as a function of time for different values of $\beta_0$.
The difference in peak flux is $15\%$ between calculations with $\beta_0=0.03$ and $\beta_0=0.2$, and the difference in the peak time is $0.35 t_0$.
Also shown is the emitted flux calculated in the free surface limit of $u(\tau=0)=0$.
In the free surface limit the flux does not depend on the value of $\beta_0$, as this parameter scales out from the equations.
But in the Eddington approximation $\mathcal{L}= f_{\rm edd} \beta_0^{-1} u(\tau=0)/(\rho_0 v_0^2) \mathcal{L}_0$ and the flux does depend on the value of $\beta_0$ (see discussion in \sref{sec:Boundary conditions}).
As the value of $\beta_0$ increases, the width of the flux pulse gets wider and its maximum gets lower, but the total emitted energy is the same.
Additionally, as the bolometric energy flux depends on the combination $f_{\rm edd} \beta_0^{-1}$, a lower value of $f_{\rm edd}$ has the same effect on the flux as a higher value of $\beta_0$.

Note that the bolometric flux does not strictly converge to the free surface limit at $\beta_0\rightarrow 0$.
This happens because as the shock velocity is lower so is the temperature lower and the relative contribution of the matter pressure becomes higher.
For example, for $\beta_0=0.03$ and $n_{\rm p,0}=10^{15}\cm^{-3}$ the matter pressure is $\approx 0.1\%$ of the total pressure at an optical depth of $\tau=\beta_0^{-1}$, while for the same density and $\beta_0=0.01$ the matter pressure is already as high as $\approx 1\%$ of the total pressure at the same optical depth.
Note also that in order to translate the presented instantaneous flux to the observed flux it is required to consider the angular dependence of the intensity in a light travel time calculation \citep[e.g.][]{Katz12}, not shown here.

\begin{figure}[h]
\includegraphics[scale=0.8]{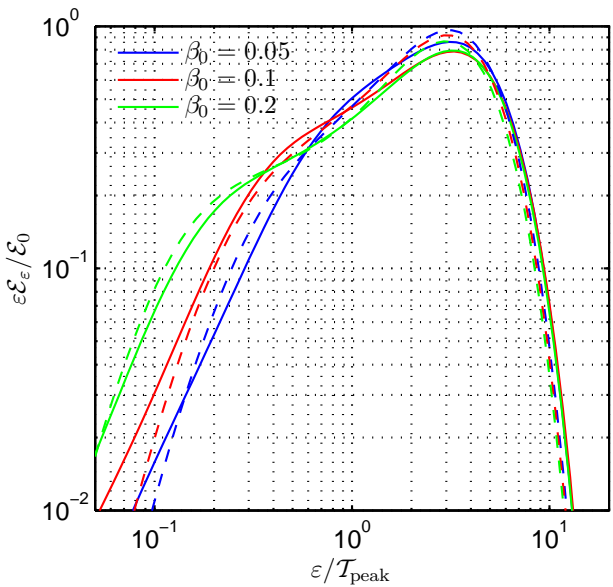}
\caption{\label{fig:Fluence_spec}
Fluence spectrum as a function of photon energy for different values of $\beta_0$. The fluence is normalized by $\mathcal{E}_0$ (eq.~\eqref{eq:e0}) and the photon energy is normalized by $\mathcal{T}_{\rm peak}$, the peak radiation temperature at the surface. Solid lines present spectral calculations for a hydrogen envelope with different values of $\beta_0$, and $n_{\rm p,0}=10^{15} \cm^{-3}$, assuming free streaming at the surface ($f_{\rm edd}=1$).
Dashed lines present calculations in the effective photon approximation.
}
\end{figure}

Figure \ref{fig:Fluence_spec} presents $\varepsilon \mathcal{E}_\varepsilon$, the spectral fluence per logarithmic unit of photon energy, in spectral calculations with different values of $\beta_0$.
The fluence is normalized by $\mathcal{E}_0$ (see eq.~\eqref{eq:e0}) and the photon energy is normalized by $\mathcal{T}_{\rm peak}$, the peak surface radiation temperature.
For comparison, also presented are calculations of the spectral fluence in the effective photon approximation, showing good agreement between the different calculations.
The presented shape of the spectral fluence is therefore confirmed here as a distinct signature of the breakout model.

Note that the fluence spectrum does not feature a high-energy tail, and it falls (almost) exponentially.
This is in contrast to the usual practice of comparing observations of XRFs to spectral models that include a power-law tail.
Unlike steady state RMSs, in the breakout problem the flow expands at the downstream.
This reduces the optical depth from which photons can diffuse from the downstream to the upstream, but also inhibits bulk Comptonization.
As can be seen, similar to steady state RMSs, bulk Comptonization does not produce a high energy power-law tail in this situation either.

\section{Discussion}\label{sec:Discussion}

The problem of a non-relativistic RMS breaking out from a stellar surface was solved numerically in the planar approximation, with detailed radiation-hydrodynamics spectral calculations.
All the main processes coupling the matter and the radiation were included in the calculations, including absorption, emission, scattering, compression and transport (see eqs.~\eqref{eq:dudt}-\eqref{eq:dedt}).
In particular, the important process of inelastic Compton scattering, which was not included in previous full radiation-hydrodynamics works, was described here with the Kompaneets equation (see eq.\eqref{eq:u_terms}).
in addition, the diffusion approximation was used to describe radiation transport, the absorption and emission processes were described with bremsstrahlung opacity and the radiation compression work term included the Doppler and abberation corrections.
We reported here a comparison of the results obtained in the effective photon approximations, presented in \citet{Sapir13}, against the exact spectral calculations performed in the present work.

A numerical scheme for solving the radiation-hydrodynamics equations was described and verified against problems with analytical spectral and bolometric solutions (see \sref{sec:Numerical scheme for the solution of the equations}).
Calculations of a steady-state RMS in a homogenous medium were presented in \sref{sec:Radiation mediated shock in a homogenous medium}, and the results were compared to the analytical hydrodynamic solutions and to the temperature profiles obtained in the effective photon approximation (see figure \ref{fig:T_vs_tau_RMS}).
The hydrodynamic profiles were in very good agreement with the analytical solutions ($<10^{-3}\%$ error).
The effective photon approximation is accurate to $3\%$ in peak temperature at shock velocity $\beta_0=v_0/c=0.05$ and has a growing discrepancy with increasing $\beta_0$.
However, even at $\beta_0=0.2$ the discrepancy is still only $\approx 27\%$.
The spectra at different points across the shock front show the signature of thermal Comptonization, supporting the effective photon approximation.
Specifically, the spectra do not feature extended high energy power-law tails, that are the signature of bulk Comptonization (see figure \ref{fig:RMS_spectrum}).

Calculations of an RMS breakout from the edge of a gas with a power-law density profile, $\rho\propto x^n$, where $x$ is the distance from the surface were presented in \sref{sec:Planar shock breakout from a stellar surface}.
This problem is defined by two parameters, $\beta_0$, the shock velocity, and $\rho_0$, the pre-shock density, and the matter composition.
The temperature profiles obtained in the calculations were compared to the ones obtained in the effective photon approximation, showing good agreement (see figure \ref{fig:T_vs_tau}).
A comparison of the peak surface temperature between the spectral calculations and the effective photon approximation was provided for different values of the breakout parameters $\beta_0$ and $\rho_0$, showing good agreement between the current and previous results.
At high shock velocities, $\beta_0=0.2$, and low densities, $n_{\rm p,0}=10^{13} \cm^{-3}$, the discrepancy between the different calculations is only $30\%$.

The bolometric energy flux emitted from the surface was calculated in the Eddington approximation relating between the flux and the energy density at the surface.
Calculations with a free streaming condition at the surface for different values of $\beta_0$, were compared to the free surface limit of $u(\tau)=0$ (see figure \ref{fig:Bol_luminosity}).
The results do not strictly converge at $\beta_0 \rightarrow 0$, but the sensitivity is very low between different values of $\beta_0$, less than a $10\%$ difference in peak flux between $\beta_0=0.01$ and $\beta_0=0.2$.

The spectral fluence, the spectral energy flux integrated over time, was compared to the universal shape obtained in \cite{Sapir13} (see figure \ref{fig:Fluence_spec}).
This observable is insensitive to the envelope density slope, to the angular distribution of the intensity at the surface, to the light travel time calculation and to mild asphericity of the emerging shock, and therefore presents a unique and robust signature of RMS breakout from a stellar envelope \citep{Sapir13}.
The spectral fluence shape calculated in the current work agrees with the spectral fluence previously calculated in the effective photon approximation, reconfirming this prediction.
Notably, the spectrum lacks a power-law high-energy tail, a feature that is usually included in comparisons to X-ray observations.

The confirmation of the spectral fluence shape by detailed calculations further justifies the statement that the spectral fluence shape is not sensitive to the shape of the instantaneous spectral flux but rather it is determined by the time-dependence of the flux and the radiation temperature \citep{Sapir13}.
In particular, an instantaneous spectral flux that is described with a characteristic photon energy indicates that the spectral fluence is also described by a characteristic photon energy, as opposed to being described by an extended power-law at high photon energies.
On the other hand, even if the (instantaneous) emission is characterized by a black-body spectrum, a simple black-body fit is bound to disagree with a fluence that is integrated over a time dependent emission, where both the flux and the radiation temperature change with time.
This also means that the non-thermal nature of a transient can not be established on the basis of the spectral fluence shape alone, if this features a characteristic photon energy.

The confirmation of the validity of the results of the effective photon approximation also imply that the discrepancies of the breakout model with respect to observations remain unresolved.
As was shown in \citet{Sapir13}, comparison of the breakout model with the breakout candidate XRF080109/SN2008D shows some discrepancies, particularly the presence of high-energy photons ($>8 \keV$) in the observed spectrum.
An additional issue is the question of missing X-ray breakout detections from SNe similar to SN1987A, posed in \citet{Sapir13}.
The detailed calculations presented here provide a similar estimate for the total number of photons emitted in SN1987A-like SNe and their characteristic energy, and there are therefore expected to be many breakout detections by past and active X-ray telescopes, but none are actually reported.

\acknowledgments
We thank Eli Waxman, Yonatan Elbaz, Boaz Katz and Peter Szabo for valuable discussions.
N.S. is partially supported by UPBC and GIF grants.

\begin{appendices}
\numberwithin{equation}{section}
\renewcommand{\thesection}{\Alph{section}}
\renewcommand{\thesubsection}{\Roman{subsection}}
\renewcommand{\theequation}{\thesection\arabic{equation}}

\section*{Appendix}

\section{Properties of the radiation temperature}\label{sec:Properties of the radiation temperature}

For a given $u_\varepsilon$, the spectral energy density in units of energy density per unit photon energy, $T_\gamma$, the radiation temperature, can be defined as an energy moment of the spectrum (see eq.~\eqref{eq:T_gamma})
\begin{equation}\label{eq:T_gamma_app}
T_\gamma=\frac{1}{4}\frac{P+Q}{u},
\end{equation}
where
\begin{equation}
\begin{split}
P=&\int_0^\infty \varepsilon u_\varepsilon d\varepsilon, \\
Q=&\int_0^\infty \frac{(hc)^3}{8\pi} \left(\frac{u_\varepsilon}{\varepsilon}\right)^2d\varepsilon, \\
u=&\int_0^\infty u_\varepsilon d\varepsilon.
\end{split}
\end{equation}
This definition holds some important properties.
The expression in eq.~\eqref{eq:T_gamma_app} for the radiation temperature is derived by integrating by parts the Kompaneets equation (the scattering term in eq.~\eqref{eq:u_terms}),
\begin{equation}\label{eq:dudt_scat_app}
\frac{\partial u_\varepsilon}{\partial t}=\rho \kappa c \frac{\varepsilon}{m_{\rm e} c^2}\frac{\partial}{\partial \varepsilon}\left[T\frac{\partial}{\partial \varepsilon}(\varepsilon u_\varepsilon)+(\varepsilon-4T)u_\varepsilon+\frac{(hc)^3}{8\pi}\left(\frac{u_\varepsilon}{\varepsilon}\right)^2\right],
\end{equation}
which yields
\begin{equation}\label{eq:emissivity_C_app}
\frac{du}{dt}=4 u (\rho \kappa c)  \frac{T-T_\gamma}{m_{\rm e} c^2}.
\end{equation}
In steady-state, there is no net energy exchange between the matter and the radiation and so the radiation temperature is equal to the matter temperature.
As the steady-state solution for the Kompaneets equation is a Bose-Einstein spectrum, for such a spectrum the radiation temperature defined by eq.~\eqref{eq:T_gamma_app} is equal to the matter temperature.

A more direct way to see this is to consider a Bose-Einstein spectrum, given by
\begin{equation}\label{eq:BE_gas}
u_\varepsilon=\frac{8\pi}{(hc)^3}\frac{\varepsilon^3}{e^{(\varepsilon+\mu)/T}-1},
\end{equation}
where $\mu$ is the chemical potential.
This spectrum is the solution for both LTE ($\mu=0$) and Compton equilibrium ($\mu>0$).
Deriving the different terms in $T_\gamma$ from this spectrum, we get
\begin{equation}
\begin{split}
P=&24 T^5  \frac{8\pi}{(hc)^3} \sum_{n=1}^\infty \frac{e^{-n\mu/T}}{n^5} , \\
Q=&24 T^5  \frac{8\pi}{(hc)^3} \sum_{n=2}^\infty e^{-n\mu/T}\left(\frac{1}{n^4}-\frac{1}{n^5}\right), \\
u=&6 T^4 \frac{8\pi}{(hc)^3} \sum_{n=1}^\infty \frac{e^{-n\mu/T}}{n^4} ,
\end{split}
\end{equation}
and using eq.~\eqref{eq:T_gamma_app} we get $T_\gamma=T$ for any given $\mu$.
Therefore $T_\gamma$ represents the correct temperature for both LTE and Compton equilibrium.

Another interesting property of $T_\gamma$ is its behavior under adiabatic compression/expansion.
For pure adiabatic compression/expansion, the spectral energy density evolves according to (see eq.~\eqref{eq:u_terms})
\begin{equation}\label{eq:dudt_comp_app}
\frac{\partial u_\varepsilon}{\partial t}=\left[\frac{4}{3}u_\varepsilon-\frac{1}{3}\frac{\partial (\varepsilon u_\varepsilon)}{\partial \varepsilon}\right]\frac{d \ln \rho}{dt},
\end{equation}
and accordingly the different terms in $T_\gamma$ evolve as
\begin{equation}
\begin{split}
\frac{dP}{dt}=&\frac{5}{3}P\frac{d \ln \rho}{dt}, \\
\frac{dQ}{dt}=&\frac{5}{3}Q\frac{d \ln \rho}{dt}, \\
\frac{du}{dt}=&\frac{4}{3}u\frac{d \ln \rho}{dt}.
\end{split}
\end{equation}
Using the chain rule for the derivative
\begin{equation}
\frac{dT_\gamma}{dt}=\frac{1}{4}\left[\frac{1}{u}\left(\frac{dP}{dt}+\frac{dQ}{dt}\right)-\frac{P+Q}{u^2}\frac{du}{dt}\right],
\end{equation}
we get
\begin{equation}
\frac{dT_\gamma}{dt}=\frac{1}{3}T_\gamma \frac{d \ln \rho}{dt},
\end{equation}
which solution is $T_\gamma \propto \rho^{1/3}$.
Therefore $T_\gamma$ behaves the same as the temperature of an ideal gas with adiabatic index $4/3$ in adiabatic compression/expansion, for any given spectrum.

An additional property of $T_\gamma$ is its relation to the thermodynamic identity $\theta=dU/dS|_V$, where $U=uV$, $V$ is the volume and $S$ is the entropy.
This relation can be used to describe the temperature also out of equilibrium \citep[e.g.][]{Casas03}.
For a gas of photons the entropy density per unit photon energy is given by \citep{Landau80,Caflisch86}
\begin{equation}\label{eq:spectral_entropy_density}
\begin{split}
s_\varepsilon=\frac{8\pi}{(hc)^3}\varepsilon^2\bigg[ & \left(1+\frac{(hc)^3}{8\pi}\frac{u_\varepsilon}{\varepsilon^3}\right)
\log\left(1+\frac{(hc)^3}{8\pi}\frac{u_\varepsilon}{\varepsilon^3}\right) \\
& -\frac{(hc)^3}{8\pi}\frac{u_\varepsilon}{\varepsilon^3}\log\left(\frac{(hc)^3}{8\pi}
\frac{u_\varepsilon}{\varepsilon^3}\right)\bigg].
\end{split}
\end{equation}
For instance, the entropy density $s=\int_0^{\infty}s_\varepsilon d\varepsilon$ for a Bose-Einstein spectrum (eq.~\eqref{eq:BE_gas}) can be evaluated with integration by parts of eq.~\eqref{eq:spectral_entropy_density}, and is
\begin{equation}
\begin{split}
s=\frac{8\pi}{(hc)^3}T^3\bigg[ & 8\sum_{n=1}^{\infty}\frac{e^{-n\mu/T}}{n^4}+2\frac{\mu}{T}\sum_{n=1}^{\infty}\frac{e^{-n\mu/T}}{n^3}
+\frac{1}{3}\left(\frac{\mu}{T}\right)^3\sum_{n=1}^{\infty}\frac{e^{-n\mu/T}}{n} \\
&-\frac{1}{3}\left(\frac{\mu}{T}\right)^4\bigg]+
\frac{8\pi}{(hc)^3}\frac{\mu^3}{3}\log\left(e^{\mu/T}-1\right),
\end{split}
\end{equation}
which converges at $\mu=0$ to the correct value of black-body radiation.
The expression in eq.~\eqref{eq:spectral_entropy_density} for the spectral entropy density also holds the correct property of zero entropy change in adiabatic compression/expansion for any given spectrum.
Using the rate of entropy density change
\begin{equation}\label{eq:dsdt_app}
\frac{ds}{dt}=\int_0^\infty\frac{1}{\varepsilon}\left[\log\left(1+\frac{(hc)^3}{8\pi}\frac{u_\varepsilon}{\varepsilon^3}\right)-
\log\left(\frac{(hc)^3}{8\pi}\frac{u_\varepsilon}{\varepsilon^3}\right)\right]\frac{\partial u_\varepsilon}{\partial t}d\varepsilon,
\end{equation}
the entropy differential $dS=sdV+VdS$, and eq.~\eqref{eq:dudt_comp_app} we get
\begin{equation}
\begin{split}
\frac{dS}{dt}=\frac{dV}{dt}\int_0^\infty\bigg[&\left(\varepsilon^2\frac{8\pi}{(hc)^3}+\frac{1}{3}\frac{\partial u_\varepsilon}{\partial \varepsilon}\right)\log\left(1+\frac{(hc)^3}{8\pi}\frac{u_\varepsilon}{\varepsilon^3}\right) \\
&-\frac{1}{3}\frac{\partial u_\varepsilon}{\partial \varepsilon}\log\left(\frac{(hc)^3}{8\pi}\frac{u_\varepsilon}{\varepsilon^3}\right)\bigg]d\varepsilon=0,
\end{split}
\end{equation}
so entropy is conserved in adiabatic compression/expansion for any given spectrum for which $u_\varepsilon(\varepsilon\rightarrow 0)=0$ and $u_\varepsilon(\varepsilon\rightarrow \infty)=0$.

Now let us consider a homogeneous radiation gas scattering with a cold electron gas with $T=0$.
Substituting eq.~\eqref{eq:dudt_scat_app} in eq.~\eqref{eq:dsdt_app} for the entropy density change and again integrating by parts we get
\begin{equation}\label{eq:dsdt_scat_app}
\frac{ds}{dt}=-4u\frac{\rho \kappa c}{m_{\rm e} c^2}.
\end{equation}
Although this negative change in entropy may seem unphysical, it is the correct description of the problem.
For such a situation the radiation is driven toward a Bose-Einstein condensation, which has zero entropy, but the decrease in entropy is balanced by the increase in entropy needed to keep the electrons at zero temperature.

Using eqs.~\eqref{eq:emissivity_C_app} and \eqref{eq:dsdt_scat_app} the thermodynamic relation $\theta=dU/dS|_V$ can be estimated, and we get $\theta=T_\gamma$.
So in the case of pure scattering over a cold electron gas the radiation temperature out of equilibrium can be defined as an internal property of the radiation spectrum alone.

\section{Test problems for the numerical code}\label{sec:Test problem for the numerical code}

The numerical scheme described in \sref{sec:Numerical scheme for the solution of the equations} was implemented and verified against test problems that involve one or more of the described radiation mechanisms and that have exact analytical solutions.
The following test problems are described in this appendix:
\begin{enumerate}[I]
\item A finite slab of homogenous plasma illuminated with a constant energy flux, without radiative interactions,
\item Release of radiation as a point energy source in a finite size box with a homogenous matter, without radiative interactions.
\item Homogenous matter and radiation interacting only through absorption and emission,
\item Homogenous matter and radiation interacting only through scattering,
\item Adiabatic expansion of a isotropic radiation gas, without radiative interactions.
\end{enumerate}

\subsection{Constant flux boundary condition}
\begin{figure}[h]
\includegraphics[scale=0.8]{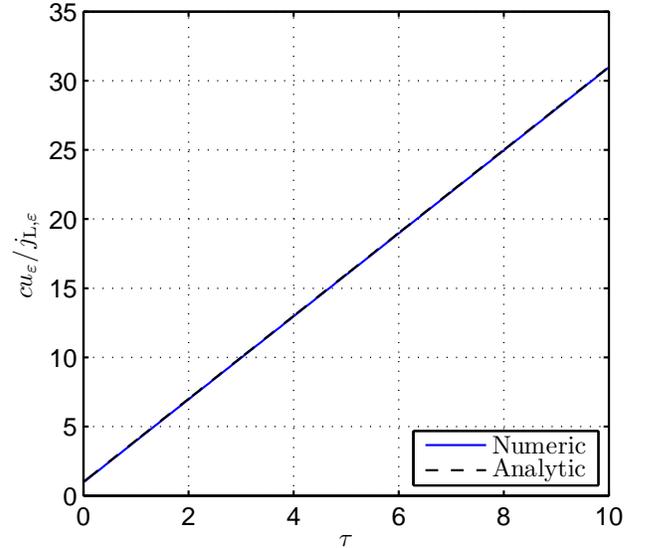}
\caption{\label{fig:u_benchmarkcurrent}
The steady-state spectral energy density as a function of optical depth in the constant energy flux problem. Blue solid line presents the numeric calculation and black dashed line presents the analytic solution according to eq.~\eqref{eq:u_current_app}.
The parameters in the calculation are $\rho_0=10^{-9} \gr\cm^{-3}$, $\kappa_\varepsilon^*=\kappa$ and $X=10/(\rho \kappa) \cm$. The time presented is $t=10 X^2 (\rho \kappa)/c$.
}
\end{figure}

This problem tests the radiation transport part of the code in the diffusion approximation.
In this problem the radiation energy equation, eq.~\eqref{eq:dudt}, is solved with only the radiation transport term included.
Consider a finite slab with size $X$ of a uniform plasma with density $\rho$ and diffusion opacity $\kappa_\varepsilon^*$, illuminated by a constant radiation beam, neglecting hydrodynamics and energy exchange between the radiation and the matter.
To describe this case, a spectral energy flux boundary condition that is constant in time is imposed at the left side, $j_{{\rm L},\varepsilon}$, and a free streaming boundary condition is imposed at the right side ($f_{\rm edd}=1$).
If the only mechanism coupling the matter to the radiation is diffusion, after some time a steady state would be achieved, with a uniform spectral energy flux over the slab.
This is basically a monochromatic problem, and each photon energy can be solved separately.
The steady state solution for the spectral energy density as a function of $\tau=\int_x^X \kappa_\varepsilon^*\rho dx$, the optical depth from the right boundary, is
\begin{equation}\label{eq:u_current_app}
u_\varepsilon=\frac{j_{{\rm L},\varepsilon}}{c}\left(f_{\rm edd}^{-1}+3\tau \right).
\end{equation}
Figure \ref{fig:u_benchmarkcurrent} presents the spectral energy density as a function of the optical depth from the surface, for a single photon energy.
Presented are the analytical solution and the numerical solution for the parameters $\rho=10^{-9} \gr\cm^{-3}$ and $\kappa_\varepsilon^*=\kappa$, with a slab of size $X=10/(\rho \kappa)$ divided into $N=100$ spatial cells, after a time $10 X^2 (\rho \kappa)/c$.
As the spectral energy density scales with $j_{{\rm L},\varepsilon}$, this parameter was chosen arbitrarily.

\subsection{Release of a point radiation source}
\begin{figure}[h]
\includegraphics[scale=0.8]{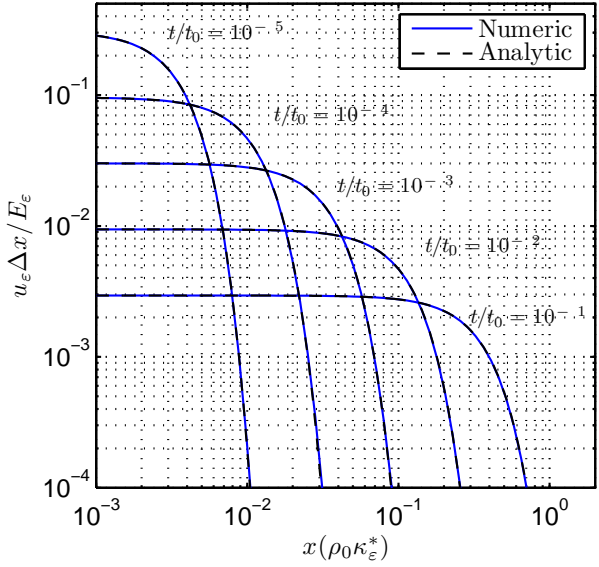}
\caption{\label{fig:u_benchmarkpoint}
Spectral energy density as a function of optical depth at different times for the point radiation source problem. Blue solid lines present the numeric calculation and the black dashed lines present the analytic solution according to eq.~\eqref{eq:u_point_app}.
The parameters in the calculation are $\rho_0=10^{-9} \gr\cm^{-3}$ and $X=10 x_0$, with $x_0=(\rho_0\kappa)^{-1}$.
The text over the curves indicates the presented times with respect to $t_0=x_0/c$.
}
\end{figure}

This problem again tests the radiation transport part of the code in the diffusion approximation.
In this problem the radiation energy equation, eq.~\eqref{eq:dudt}, is solved with only the radiation transport term included.
Consider a finite slab with size $X$ of a uniform plasma with density $\rho$ and diffusion opacity $\kappa_\varepsilon^*$, where a point radiation source of energy per unit area per unit photon energy $E_\varepsilon$, is instantaneously released at time $t=0$ at $x=0$, with a reflective boundary condition imposed on the left side, $j_{\rm L,\varepsilon}=0$, and a zero radiation energy boundary condition imposed at the right side, $u_\varepsilon(x=X)=0$.
Again, this is basically a monochromatic problem, and each photon energy can be solved separately.
The solution for the spectral energy density as a function of spatial position and time is given by
\begin{equation}\label{eq:u_point_app}
u_\varepsilon=\frac{2E_\varepsilon}{\Gamma(1/2)}\left(\frac{3}{4}\frac{\rho \kappa_\varepsilon^*}{ct}\right)^{1/2}e^{-3\rho \kappa_\varepsilon^* x^2/4ct}.
\end{equation}
Figure \ref{fig:u_benchmarkpoint} presents the spectral energy density as a function of the spatial position, for a single photon energy, at several times.
Presented are the analytical solution and the numerical solution for the parameters $\rho=10^{-9} \gr\cm^{-3}$ and $\kappa_\varepsilon^*=\kappa$, with a slab of size $X=10/(\rho \kappa)$ divided into $N=100$ spatial cells.
As the spectral energy density scales with $E_\varepsilon$, this parameter was chosen arbitrarily.

\subsection{Matter and radiation interacting through pure absorption and emission}
\begin{figure}[h]
\includegraphics[scale=0.8]{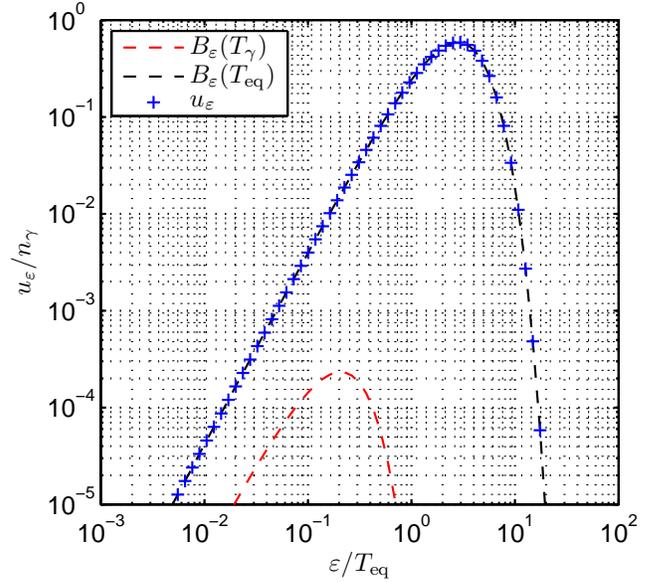}
\caption{\label{fig:u_spec_benchmarkemis}
Spectral energy density as a function of photon energy in the pure emission and absorption test problem. Presented are the initial black-body spectrum (red dashed line), the calculated steady-state spectrum (blue crosses) and the analytical steady-state spectrum, from eqs.~\eqref{eq:Planck_spectrum} and \eqref{eq:app_Teq_emis} (dashed black line). Presented is a calculation for the parameters $n_{\rm p,0}=10^{15} \cm^{-3}$, $T=1 \keV$ and $T_\gamma=1 \eV$, after time $50t_{\rm abs}$ (see eq.~\eqref{eq:app_tabs}). The spectra are normalized by $n_\gamma=a_{\rm BB}T_{\rm eq}^3/2.7$.
}
\end{figure}

This problem tests the absorption and emission part of the code.
In this problem the energy equations, eqs.~\eqref{eq:dudt} and \eqref{eq:dedt}, are solved with only the absorption and emission terms included.
Consider an infinite homogeneous medium filled initially with a hydrogen plasma of proton number density and temperature $n_{\rm p}=\rho_0/m_{\rm p}$ and $T$, respectively, and a black-body radiation with $T_\gamma<T$.
As the matter continues to produce photons, the matter and radiation will exchange energy with each other until they equilibrate temperatures.
In steady-state, the radiation will achieve a black-body spectrum, eq.~\eqref{eq:Planck_spectrum}, where $T_{\rm eq}$, the equilibrium temperature, is determined by energy conservation,
\begin{equation}\label{eq:app_Teq_emis}
a_{\rm BB} T_{\rm eq}^4+3n_{\rm p}T_{\rm eq}=a_{\rm BB} T_\gamma^4+3n_{\rm p}T.
\end{equation}
In the limit $T_{\rm eq} \ll T$, the time it takes for the matter to achieve its steady-state temperature is roughly $t_{\rm cool}=3n_{\rm p}T/j_{\rm B}(n_{\rm p},T)$, with the bremsstrahlung emissivity given by eq.~\eqref{eq:emissivity_B}.
However, it takes a much longer time for the radiation to achieve a black-body spectrum, as high energy photons are slowly absorbed due to the decrease of opacity with photon energy.
This time scale is roughly
\begin{equation}\label{eq:app_tabs}
t_{\rm abs}=\left[\rho_0 \kappa_\varepsilon(\rho_0,T_{\rm eq},\varepsilon=T) c\right]^{-1}.
\end{equation}
Figure \ref{fig:u_spec_benchmarkemis} presents the calculated and analytic equilibrium spectral energy density for the parameters $n_{\rm p,0}=10^{15} \cm^{-3}$, $T=1 \keV$ and $T_\gamma=1 \eV$, with a spectral grid of $100$ points divided logarithmically over photon energies from $0.01 T_\gamma$ to $100 T$.
As can be seen, the calculated spectrum agrees with the theoretical derivation.

\subsection{Matter and radiation interacting through pure scattering}
\begin{figure}[h]
\includegraphics[scale=0.8]{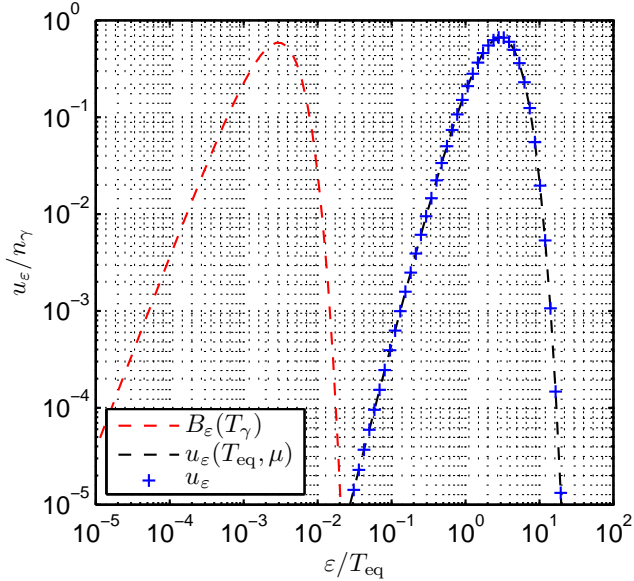}
\caption{\label{fig:u_spec_benchmarkscat}
Spectral energy density as a function of photon energy in the pure scattering test problem. Presented are the initial black-body spectrum (red dashed line), the calculated steady-state spectrum (blue crosses) and the analytic steady-state spectrum, from eqs.~\eqref{eq:BE_gas}, \eqref{eq:app_Teq} and \eqref{eq:app_mu} (dashed black line). Presented is a calculation for the parameters $n_{\rm p,0}=10^{15} \cm^{-3}$, $T=1 \keV$ and $T_\gamma=1 \eV$, after time $10 t_{\rm cool}$ (see eq.~\eqref{eq:app_tcool}). The spectra are normalized by $n_\gamma$ (eq.~\eqref{eq:app_n_ga}).
}
\end{figure}

This problem tests the scattering part of the code.
In this problem the energy equations, eqs.~\eqref{eq:dudt} and \eqref{eq:dedt}, are solved with only the scattering term included.
Consider an infinite homogeneous medium filled initially with a hydrogen plasma of number density and temperature $n_{\rm p}$ and $T$, respectively, and a black-body radiation with $T_\gamma<T$.
If Compton scattering is the only process coupling the radiation to the matter, the number of photons is conserved and the matter and radiation will exchange energy with each other until they equilibrate temperatures.
In steady-state, the radiation will achieve a Bose-Einstein spectrum, eq.~\eqref{eq:BE_gas}, where $\mu$, the chemical potential, is determined by photon number conservation, and $T_{\rm eq}$, the equilibrium temperature, is determined by energy conservation,
\begin{equation}
\begin{split}
16\pi \left(\frac{T_{\rm eq}}{hc}\right)^3 \sum_{n=1}^\infty \frac{e^{-n\mu/T_{\rm eq}}}{n^3}&=n_\gamma, \\
T_{\rm eq}\left(\frac{\sum_{n=1}^\infty n^{-4}e^{-n\mu/T_{\rm eq}}}{\sum_{n=1}^\infty n^{-3}e^{-n\mu/T_{\rm eq}}}n_\gamma+n_{\rm p}\right)&=\frac{\zeta(4)}{\zeta(3)}n_\gamma T_\gamma+n_{\rm p} T,
\end{split}
\end{equation}
where the (conserved) photon number density is
\begin{equation}\label{eq:app_n_ga}
n_\gamma=16\pi \left(\frac{T_\gamma}{hc}\right)^3\zeta(3),
\end{equation}
and $\zeta(x)$ is the Riemann zeta function.
In the limit $T_{\rm eq} \gg T_\gamma$, we get $\mu\gg T_{\rm eq}$, in that case the equilibrium temperature is given by
\begin{equation}\label{eq:app_Teq}
T_{\rm eq}=\frac{\left[\zeta(4)/\zeta(3)\right]n_\gamma T_\gamma+n_{\rm p} T}{n_\gamma+n_{\rm p}},
\end{equation}
and the chemical potential is
\begin{equation}\label{eq:app_mu}
\mu=T_{\rm eq}\left[3\log(T_{\rm eq}/T_\gamma)-\log(\zeta(3))\right].
\end{equation}
The time it takes to achieve this steady-state is roughly
\begin{equation}\label{eq:app_tcool}
t_{\rm cool}= \frac{m_{\rm e}c^2}{4 T_{\rm eq}}\log\left(\frac{T_{\rm eq}}{T_\gamma}\right)(n_{\rm p} \sigT c)^{-1}.
\end{equation}
Figure \ref{fig:u_spec_benchmarkscat} presents the calculated and analytic equilibrium spectral energy density for the parameters $n_{\rm p,0}=10^{15} \cm^{-3}$, $T=1 \keV$ and $T_\gamma=1 \eV$, with a spectral grid of $100$ points divided logarithmically over photon energies from $0.01 T_\gamma$ to $100 T$.
As can be seen, the calculated spectrum agrees with the theoretical derivation.

\subsection{Expansion of a radiation gas}
\begin{figure}[h]
\includegraphics[scale=0.8]{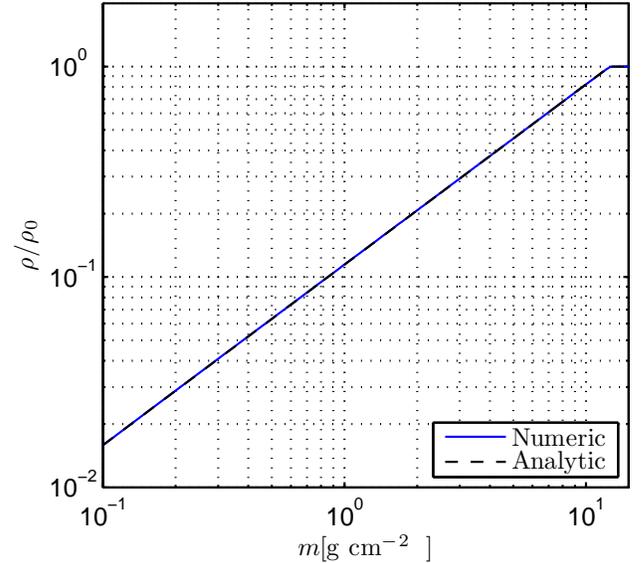}
\caption{\label{fig:rho_benchmarkexp}
Density as a function of the mass coordinate in the radiation gas expansion test problem. Presented are the calculated density (blue solid line) and the analytic density from eq.~\eqref{eq:app_rho_exp} (black dashed line).
The parameters in the calculation are $\rho_0=10^{-9} \gr\cm^{-3}$, $T_0=0.1 \keV$, $X=10^{10} \cm$, and the time presented is  $t=(3/4)X/c_0$.
}
\end{figure}

\begin{figure}[h]
\includegraphics[scale=0.8]{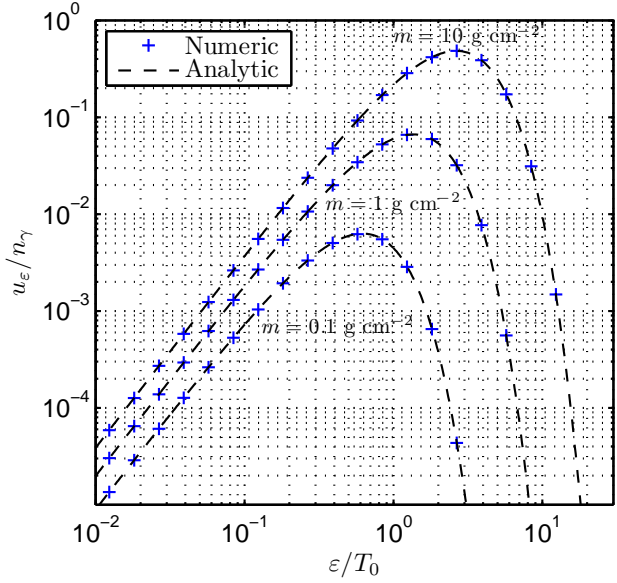}
\caption{\label{fig:u_spec_benchmarkexp}
Spectral energy density as a function of photon energy for different mass elements in the radiation gas expansion test problem. Presented are the calculated spectra (blue crosses) and the black-body spectra of different mass elements (black dashed line), using eq.~\eqref{eq:app_rho_exp}. The parameters in the calculation are the same as in figure \ref{fig:rho_benchmarkexp}, and the spectra are normalized by $n_\gamma$ (see eq.~\eqref{eq:app_n_ga}). The text denotes the mass element from which the spectrum was taken.
}
\end{figure}

This problem tests the compression part of the code for the radiation gas alone.
In this problem the flow equations and the radiation energy equation, eqs.~\eqref{eq:continuity_equation}-\eqref{eq:dudt}, are solved without the radiative interactions with the matter and without radiation transport.
Consider a finite slab with size $X$ of a uniform matter with density $\rho_0$ and zero temperature, with a uniform black-body radiation of temperature $T_0$.
After releasing this gas at time $t=0$, a rarefaction wave propagates inside the matter at the sound speed
\begin{equation}
c_0=\left(\frac{4}{9}\frac{u_0}{\rho_0}\right)^{1/2},
\end{equation}
where $u_0=a_{\rm BB}T_0^4$.
Using the mass coordinate per unit area $m=\int_x^X \rho dx$, the density profile at time $t$ is given by
\begin{equation}\label{eq:app_rho_exp}
\rho(m,t)=
\begin{cases}
\rho_0^{1/7}\left(\frac{m}{c_0 t}\right)^{6/7}&, \quad m < \rho_0 c_0 t,\\
\rho_0&, \quad m > \rho_0 c_0 t,
\end{cases}
\end{equation}
the velocity is $v=6c_0[1-(\rho/\rho_0)^{1/6}]$, the bolometric energy density is $u=u_0(\rho/\rho_0)^{4/3}$, the radiation temperature is $T=T_0(\rho/\rho_0)^{1/3}$, and the spectrum remains black-body everywhere.
Figure \ref{fig:rho_benchmarkexp} presents the density profile for a calculation with the parameters $\rho_0=10^{-9} \gr\cm^{-3}$, $T_0=0.1 \keV$ and a slab of size $X=10^{10} \cm$ divided into $400$ equal size cells, and a spectral grid of $30$ points divided logarithmically over photon energies from $10^{-3} T_0$ to $100 T_0$.
The presented profile is at time $t=(3/4)X/c_0$.
Figure \ref{fig:u_spec_benchmarkexp} presents the spectral energy density at different mass elements for the same calculation at the same time.
As can be seen, the black-body spectral shape is indeed preserved in the calculation, with the correct temperature.


\section{Structure of a radiation mediated shock}\label{sec:Structure of a radiation mediated shock}

The hydrodynamic profiles describing the RMS's structure have been solved analytically \citep{Weaver76}.
We provide here the results of this derivation for the hydrodynamic variables in terms of the mass coordinate (per unit area).
For a shock moving in the positive direction of mass, with an upstream density of $\rho_0$, shock velocity of $v_0$, and opacity $\kappa$, the density, pressure, velocity, energy flux and position of the flow as a function of the mass coordinate are given by
\begin{equation}\label{eq:RMS_structure}
\begin{split}
\rho_{\rm W}(m)&=\rho_0\frac{e^{3m/m_0}+1}{e^{3m/m_0}+1/7}, \\
v_{\rm W}(m)&=v_0 \left(1-\frac{\rho_0}{\rho_{\rm W}(m)}\right), \\
p_{\rm W}(m)&=\rho_0 v_0^2 \left(1-\frac{\rho_0}{\rho_{\rm W}(m)}\right), \\
j_{\rm W}(m)&=\rho_0 v_0^3 \left[-\frac{1}{2}+4\frac{\rho_0}{\rho_{\rm W}(m)}-\frac{7}{2}\left(\frac{\rho_0}{\rho_{\rm W}(m)}\right)^2\right], \\
x_{\rm W}(m)&=\frac{1}{7}\left[\frac{m}{\rho_0}+2\ln\left(e^{3m/m_0}+1\right)\frac{m_0}{\rho_0}\right]-\frac{2}{7}\ln(2)\frac{m_0}{\rho_0},
\end{split}
\end{equation}
where $m_0=(\beta_0\kappa)^{-1}$, and $m=0$ is the point where the energy flux is at its maximum, denoting the shock's position.
Two other functions provide the spatial position at the far downstream and far upstream, from which the spatial deformation with respect to a discontinuous shock front can be calculated,
\begin{equation}\label{eq:x_deformation}
\begin{split}
x_{\rm W,m \rightarrow -\infty}(m) =&\frac{1}{7}\frac{m}{\rho_0}-\frac{2}{7}\ln(2)\frac{m_0}{\rho_0}, \\
x_{\rm W,m \rightarrow \infty}(m) =&\frac{m}{\rho_0}-\frac{2}{7}\ln(2)\frac{m_0}{\rho_0}.
\end{split}
\end{equation}

\section{Hydrodynamic profiles in the self-similar solution}\label{sec:Hydrodynamic profiles in the self-similar solution}
The problem of a shock wave propagating in an ideal gas with a power-law density profile ending at an edge, $\rho_{\rm init}=\tilde{b} x^n$, has been solved by a self-similar analysis \citep{Gandel'Man56,Sakurai60}.
In order to describe the hydrodynamic profiles of the self-similar solution we follow \citet{Zeldovich67} and \citet{Matzner99}.

Denoting the shock's position as $x_{\rm sh}(t)$, where $t=0$ is taken to be the time when the shock emerges at the surface, the similarity relation for the shock velocity is assumed as $v_{\rm sh} =-\tilde{a} x_{\rm sh}^{-\lambda}$.
In terms of the parameters in this paper, the proportionality factors for the density and the shock velocity are parameterized as
\begin{equation}
\tilde{b}=\rho_0\left(\frac{\rho_0\kappa \beta_0}{n+1}\right)^n, \quad \tilde{a}= \beta_0 c \left(\frac{\rho_0\kappa \beta_0}{n+1}\right)^{-\lambda}.
\end{equation}
Denoting the mass coordinate (per unit area) of the shock as $m_{\rm sh}$, the shock's position is given by
\begin{equation}\label{eq:xsh_app}
x_{\rm sh}=\frac{n+1}{\rho_0\kappa \beta_0}\left(\frac{m_{\rm sh}}{m_0}\right)^{1/(n+1)},
\end{equation}
where $m_0=(\kappa\beta_0)^{-1}$, and the shock velocity is
\begin{equation}
v_{\rm sh}=-v_0\left(\frac{m_{\rm sh}}{m_0}\right)^{-\lambda/(n+1)}.
\end{equation}
Through the self-similar transformation (in Eulerian coordinates),
\begin{equation}
\begin{split}
& \xi(x,t)=x/x_{\rm sh}(t), \quad v(x,t)=v_{\rm sh} \xi U(\xi), \\
& c(x,t)=v_{\rm sh} \xi C(\xi), \quad \rho(x,t)=\tilde{b}x_{\rm sh}^n G(\xi),
\end{split}
\end{equation}
where $c$ is the sound speed, related to the pressure by $p=\rho c^2/\gamma$ and $\gamma$ is the adiabatic index, the gas hydrodynamic equations (without diffusion) can be reduced to two quadratures describing $\xi(U)$ and $\xi(C)$ and a differential equation describing the $U(C)$ curve,
\begin{equation}\label{eq:dUdC}
\frac{dU}{dC}=\frac{\Delta_1(U,C)}{\Delta_2(U,C)},
\end{equation}
where
\begin{equation}
\begin{split}
\Delta_1 &= U(1-U)(1-U+\lambda)-C^2\left(U+\frac{n-2\lambda}{\gamma}\right), \\
\Delta_2 &= C\left[(1-U)(1-U+\lambda)+\frac{\gamma-1}{2}\lambda U-C^2-\frac{2\lambda+n(\gamma-1)}{2\gamma}\frac{C^2}{1-U}\right].
\end{split}
\end{equation}

The correct self-similar exponent $\lambda$ that is self-consistent with the self-similarity assumption is obtained when the integrated curve $U(C)$ intersects with the curve $U=1-C$ at the singular point $\Delta_1=\Delta_2=0$.
This singular point therefore defines a critical self-similar sound velocity,
\begin{equation}
C_{\rm c}=\frac{\lambda \gamma}{n+(\gamma-2)\lambda}.
\end{equation}
The self-similar exponent can be found by numerically integrating eq.~\eqref{eq:dUdC} from the Hugoniot boundary condition at $\xi=1$,
\begin{equation}
C(1)=\frac{\left[2\gamma(\gamma-1)\right]^{1/2}}{\gamma+1}, \quad U(1)=\frac{2}{\gamma+1},
\end{equation}
and till the critical self-similar sound velocity $C_{\rm c}$.
The self-consistent solution for $\lambda$ is the one for which $U(C_{\rm c})=1-C_{\rm c}$.

After the shock's passage, each mass element expands adiabatically, and the hydrodynamic profiles can be solved in Lagrangian coordinates.
Defining a self-similar time coordinate $\eta=t/t_{\rm sh}(m)$ and a self-similar position coordinate $S(\eta)=x(m)/x_{\rm sh}(m)$, and using the relation $t_{\rm sh}=(1+\lambda)^{-1}x_{\rm sh}/v_{\rm sh}$, the following derivative is obtained
\begin{equation}
\left.\frac{dx}{dx_{\rm sh}}\right|_{t}=S(\eta)-(1+\lambda)\eta S'(\eta).
\end{equation}
For each mass element $\rho_{\rm init}dx=\rho(x_{\rm sh},\eta)dx_{\rm sh}$, and the density of a mass element in the self-similar coordinates is
\begin{equation}
\rho_{\rm S}(x_{\rm sh},\eta)=\frac{\rho_{\rm sh}}{S(\eta)-(1+\lambda)\eta S'(\eta)},
\end{equation}
where $\rho_{\rm sh}=\rho_{\rm init}(x_{\rm sh})$.

With these definitions, the velocity of a mass element is $v_{\rm S}=S'(\eta) x_{\rm sh}/t_{\rm sh}$, and the acceleration is $a_{\rm S}=S''(\eta)x_{\rm sh}/t_{\rm sh}^2$.
But from momentum conservation the acceleration is also $a_{\rm S}=-(dp/dx)_{t}/\rho=-(dp/dx)_{t_{\rm sh}}/\rho_{\rm sh}$, and taking the evolution of the pressure as adiabatic,
\begin{equation}\label{eq:p_Sakurai}
p_{\rm S}(x_{\rm sh},\eta)=\frac{2}{\gamma+1}\left(\frac{\gamma-1}{\gamma+1}\frac{\rho_{\rm S}}{\rho_{\rm sh}}\right)^\gamma \rho_{\rm sh} v_{\rm sh}^2,
\end{equation}
the following second order differential equation is obtained:
\begin{equation}
\begin{split}
S''=&\frac{1}{(1+\lambda)^2}\left[(n-2\lambda)\left(S-(1+\lambda)\eta S'\right)-\lambda(1+\lambda)\gamma \eta S'\right] \\
&\cdot \left[\gamma \eta^2-\frac{\gamma+1}{2}\left(\frac{\gamma+1}{\gamma-1}\right)^\gamma\left(S-(1+\lambda)\eta S'\right)^{\gamma+1}\right]^{-1}.
\end{split}
\end{equation}
The flow is then completely described by the solution to $S(\eta)$, $S'(\eta)$ and $S''(\eta)$ with the Hugoniot initial conditions
\begin{equation}
S(1)=1, \quad S'(1)=\frac{2}{(\gamma+1)(1+\lambda)}.
\end{equation}
Accordingly, for a shock moving in the negative direction of mass, the density, pressure, velocity, acceleration and position of the flow as a function of the mass coordinate and time are
\begin{equation}\label{eq:Sakurai_flow}
\begin{split}
\rho_{\rm S}(m,t)&=
\begin{cases}
\rho_{\rm sh}(m)\left[S(\eta)-(1+\lambda)\eta S'(\eta)\right]^{-1}&, \quad \eta \leq 1, \\
\rho_{\rm sh}(m)&, \quad \eta > 1,
\end{cases} \\
p_{\rm S}(m,t)&=
\begin{cases}
\frac{2}{\gamma+1}\left(\frac{\gamma-1}{\gamma+1}\right)^\gamma\left[\frac{\rho_{\rm S}(m,t)}{\rho_{\rm sh}(m)}\right]^{\gamma} \rho_{\rm sh}(m) v_{\rm sh}^2(m)&, \quad \eta \leq 1, \\
0&, \quad \eta > 1,
\end{cases} \\
v_{\rm S}(m,t)&=
\begin{cases}
S'(\eta) x_{\rm sh}(m)/t_{\rm sh}(m) &, \quad \eta \leq 1, \\
0&, \quad \eta > 1,
\end{cases} \\
a_{\rm S}(m,t)&=
\begin{cases}
S''(\eta)x_{\rm sh}(m)/t_{\rm sh}^2(m)&, \quad \eta \leq 1, \\
0&, \quad \eta > 1,
\end{cases} \\
x_{\rm S}(m,t)&=
\begin{cases}
S(\eta) x_{\rm sh}(m)&, \quad \eta \leq 1, \\
x_{\rm sh}(m)&, \quad \eta > 1,
\end{cases}
\end{split}
\end{equation}
where $\eta=t/t_{\rm sh}(m)$, $x_{\rm sh}(m)=(n+1)(\rho_0\kappa \beta_0)^{-1}(m/m_0)^{1/(n+1)}$, $t_{\rm sh}(m)=-(n+1)(\rho_0\kappa \beta_0^2 c)^{-1}(1+\lambda)^{-1}(m/m_0)^{(1-\lambda)/(n+1)}$ and $\rho_{\rm sh}(m)=\rho_0(m/m_0)^{n/(n+1)}$.

\end{appendices}

\bibliographystyle{apj}

\end{document}

%% file: Definitions.tex
\newcommand{\DDir}{\relax{D\kern-.7em{/}}}

\newcommand{\be}{\begin{equation}}
\newcommand{\ee}{\end{equation}}
\newcommand{\bea}{\begin{equation*}}
\newcommand{\eea}{\end{equation*}}

\newcommand{\nin}{\relax{\in\kern-.8em{/}}}

\newcommand{\cm}{\mbox{ cm}}

\newcommand{\se}{\mbox{ s}}

\newcommand{\erg}{\mbox{ erg}}

\newcommand{\meV}{\mbox{ meV}}
\newcommand{\eV}{\mbox{ eV}}
\newcommand{\keV}{\mbox{ keV}}

\newcommand{\gr}{\mbox{ g}}

\newcommand{\Msun}{M_{\odot}}

\newcommand{\sref}{\S~\ref}